\documentclass[aps,prd,twocolumn,floatfix,nofootinbib]{revtex4}
\usepackage{graphicx}
\usepackage{amsmath}
\usepackage{latexsym}
\usepackage{here}
\usepackage{array}
\usepackage{dcolumn}% Align table columns on decimal point
\usepackage{bm}% bold math

\newcommand{\half}{{\textstyle\frac{1}{2}}}

\def\lsim{\mathrel{\rlap{\raise 2.5pt \hbox{$<$}}\lower 2.5pt\hbox{$\sim$}}}
\def\gsim{\mathrel{\rlap{\raise 2.5pt \hbox{$>$}}\lower 2.5pt\hbox{$\sim$}}}

\def\vecalpha{{\pmb\alpha}}
\def\GeV{{\text{GeV}}}

\renewcommand{\Re}{{\rm Re\thinspace}}
\renewcommand{\Im}{{\rm Im\thinspace}}

\allowdisplaybreaks %!!!!
\begin{document}
% Use the \preprint command to place your local institutional report
% number in the upper righthand corner of the title page in preprint mode.
% Multiple \preprint commands are allowed.
% Use the 'preprintnumbers' class option to override journal defaults
% to display numbers if necessary
%\preprint{}

%Title of paper
\title{Constraining the Two-Higgs-Doublet-Model parameter space
}

% repeat the \author .. \affiliation  etc. as needed
% \email, \thanks, \homepage, \altaffiliation all apply to the current
% author. Explanatory text should go in the []'s, actual e-mail
% address or url should go in the {}'s for \email and \homepage.
% Please use the appropriate macro foreach each type of information

% \affiliation command applies to all authors since the last
% \affiliation command. The \affiliation command should follow the
% other information
% \affiliation can be followed by \email, \homepage, \thanks as well.
\author{Abdul Wahab El Kaffas}
\email[]{awkaffas@ift.uib.no}
%\homepage[]{Your web page}
%\thanks{}
%\altaffiliation{}
\affiliation{Department of Physics and Technology, University of Bergen,
Postboks 7803, N-5020 Bergen, Norway}
\author{Odd Magne Ogreid}
\email[]{omo@hib.no}
%\homepage[]{Your web page}
%\thanks{}
%\altaffiliation{}
\affiliation{Bergen University College, Bergen, Norway}
\author{Per Osland}
\email[]{per.osland@ift.uib.no}
%\homepage[]{Your web page}
%\thanks{}
%\altaffiliation{}
\affiliation{Department of Physics and Technology, University of Bergen,
Postboks 7803, N-5020 Bergen, Norway}

\date{\today}

%%%%%%%%%%%%%%%%%%%%%%%%%%%%%%%%%%%%%%%%%%%%%%%%
\begin{abstract}
We confront the Two-Higgs-Doublet Model with a variety of experimental
constraints as well as theoretical consistency conditions.  The most
constraining data are the $\bar B\to X_s\gamma$ decay rate (at low values of
$M_{H^\pm}$), and $\Delta\rho$ (at both low and high $M_{H^\pm}$).  We also
take into account the $B\bar B$ oscillation rate and $R_b$, or the width
$\Gamma(Z\to b\bar b)$ (both of which restrict the model at low values of
$\tan\beta$), and the $B^-\to\tau\nu_\tau$ decay rate, which restricts the
model at high $\tan\beta$ and low $M_{H^\pm}$.  Furthermore, the LEP2
non-discovery of a light, neutral Higgs boson is considered, as well as the
muon anomalous magnetic moment. Since perturbative unitarity excludes high
values of $\tan\beta$, the model turns out to be very constrained.  We outline
the remaining allowed regions in the $\tan\beta$--$M_{H^\pm}$ plane for
different values of the masses of the two lightest neutral Higgs bosons, and
describe some of their properties.
\end{abstract}

% insert suggested PACS numbers in braces on next line
% insert suggested keywords - APS authors don't need to do this
%\keywords{}

%\maketitle must follow title, authors, abstract, \pacs, and \keywords
\maketitle

%%%%%%%%%%%%%%%%%%%%%%%%%%%%%%%%%%%%%%%%%%%%%%%%%%%%%%%%%%%%%%%%%%%%%%%%
\section{Introduction and notation}
\setcounter{equation}{0}
%%%%%%%%%%%%%%%%%%%%%%%%%%%%%%%%%%%%%%%%%%%%%%%%%%%%%%%%%%%%%%%%%%%%%%%%
As compared with the Standard Model (SM), the Two-Higgs-Doublet Model (2HDM)
allows for an additional mechanism for CP violation
\cite{Lee:1973iz,Weinberg:1976hu,Branco:1985aq,Accomando:2006ga}.  This is
welcome, in view of baryogenesis \cite{Riotto:1999yt,Dine:2003ax}, and one of
the main reasons for continued interest in the model.

Several experimental constraints restrict it.  The $B-\bar B$ oscillations and
branching ratio $R_b$ exclude low values of $\tan\beta$, whereas the $\bar
B\to X_s\gamma$ rate excludes low values of the charged-Higgs mass,
$M_{H^\pm}$.  The precise measurements at LEP of the $\rho$ parameter
constrain the mass splitting in the Higgs sector, and force the masses to be
not far from the $Z$ mass scale \cite{Bertolini:1985ia}.  These individual
constraints are all well-known, but we are not aware of any dedicated attempt
to combine them, other than those of \cite{Grant:1994ak,Cheung:2003pw}.  The
present study aims to go beyond that of \cite{Cheung:2003pw}, by using more
complete and more up-to-date experimental results, as well as more accurate
theoretical predictions for the above quantities.

From the theoretical point of view, there are also various consistency
conditions.  The potential has to be positive for large values of the fields
\cite{Deshpande:1977rw,ElKaffas:2006nt}.  We also require the tree-level
Higgs--Higgs scattering amplitudes to be unitary
\cite{Kanemura:1993hm,Akeroyd:2000wc,Ginzburg:2003fe}.  Together, these
constraints dramatically reduce the allowed parameter space of the model.

The present study is limited to the 2HDM~(II), which is defined by having one
Higgs doublet ($\Phi_2$) couple to the up-type quarks, and the other
($\Phi_1$) to the down-type quarks \cite{HHG}.

We write the general 2HDM potential as:
\begin{align}
\label{Eq:pot_5}
V&=\frac{\lambda_1}{2}(\Phi_1^\dagger\Phi_1)^2
+\frac{\lambda_2}{2}(\Phi_2^\dagger\Phi_2)^2
+\lambda_3(\Phi_1^\dagger\Phi_1) (\Phi_2^\dagger\Phi_2) \nonumber \\
&+\lambda_4(\Phi_1^\dagger\Phi_2) (\Phi_2^\dagger\Phi_1)
+\frac{1}{2}\left[\lambda_5(\Phi_1^\dagger\Phi_2)^2+{\rm h.c.}\right] \\
&-\frac{1}{2}\left\{m_{11}^2(\Phi_1^\dagger\Phi_1)
\!+\!\left[m_{12}^2 (\Phi_1^\dagger\Phi_2)\!+\!{\rm h.c.}\right]
\!+\!m_{22}^2(\Phi_2^\dagger\Phi_2)\right\} \nonumber
\end{align}
Thus, the $Z_2$ symmetry will be respected by the quartic terms, and
Flavour-Changing Neutral Currents are constrained \cite{Glashow:1976nt}.  We
shall refer to this model (without the $\lambda_6$ and $\lambda_7$ terms) as
the $\text{2HDM}_5$. The more general model, with also $\lambda_6$ and
$\lambda_7$ couplings, will be discussed elsewhere.

We allow for CP violation, i.e., $\lambda_5$ and $m_{12}^2$ may be complex.  
Thus, the neutral sector will be
governed by a $3\times3$ mixing matrix, parametrized in terms of the angles
$\alpha_1$, $\alpha_2$ and $\alpha_3$ as in
\cite{Accomando:2006ga,Khater:2003wq}:
\begin{equation} \label{Eq:R-matrix}
R=
\begin{pmatrix}
c_1\,c_2 & s_1\,c_2 & s_2 \\
- (c_1\,s_2\,s_3\!+\!s_1\,c_3) 
& c_1\,c_3\!-\!s_1\,s_2\,s_3 & c_2\,s_3 \\
- c_1\,s_2\,c_3\!+\!s_1\,s_3 
& - (c_1\,s_3\!+\!s_1\,s_2\,c_3) & c_2\,c_3
\end{pmatrix}
\end{equation}
where $c_1=\cos\alpha_1$, $s_1=\sin\alpha_1$, etc., and
\begin{equation} \label{Eq:alphas}
-\frac{\pi}{2}<\alpha_1\le\frac{\pi}{2},\quad
-\frac{\pi}{2}<\alpha_2\le\frac{\pi}{2},\quad
0\le\alpha_3\le\frac{\pi}{2}.
\end{equation}
(In ref.~\cite{Khater:2003wq}, the angles are denoted as
$\tilde\alpha=\alpha_1$, $\alpha_b=\alpha_2$, $\alpha_c=\alpha_3$.) 
For a discussion of this parameter space, including 
the CP-nonviolating limits, see \cite{Kaffas:2007rq}.
We will use the terminology
``general 2HDM'' as a reminder that CP violation is allowed. The present study
extends that of \cite{Cheung:2003pw} also in this respect.

Rather than taking the parameters of the potential (\ref{Eq:pot_5}) to
describe the model, we take the two lightest neutral Higgs boson masses, $M_1$
and $M_2$, together with the mixing angles (\ref{Eq:alphas}), the charged
Higgs boson mass, $M_{H^\pm}$, and $\tan\beta$ as our basic parameters.  (The
third neutral Higgs boson mass, $M_3$, is then a derived quantity.)

It is convenient to split the constraints into three categories:
\begin{itemize}
\item[(i)] Theoretical consistency constraints:
positivity of the potential \cite{Deshpande:1977rw,ElKaffas:2006nt} and
perturbative unitarity \cite{Kanemura:1993hm,Akeroyd:2000wc,Ginzburg:2003fe},
\item[(ii)] Experimental constraints on the charged-Higgs sector.
These all come from $B$-physics, and are due to $b\to s\gamma$,
$B$--$\bar B$ oscillations, and $B\to \tau\nu_\tau$.
They are all independent of the neutral sector.
\item[(iii)] Experimental constraints on the neutral sector.  These are
predominantly due to the precise measurements of $R_b$, non-observation of a
neutral Higgs boson at LEP2, $\Delta\rho$, and $a_\mu=\half(g-2)_\mu$.
\end{itemize}
The first and third categories of constraints will depend on the neutral
sector, i.e., the neutral Higgs masses and the mixing matrix.  The second
category is due to physical effects of the charged-Higgs Yukawa coupling in
the $B$-physics sector.  These are ``general'' in the sense that they do not
depend on the spectrum of neutral Higgs bosons, i.e., they do not depend on
the mixing (and possible CP violation) in the neutral sector.

When considering the different experimental constraints, our basic approach
will be that they are all in agreement with the Standard Model, and simply let
the experimental or theoretical uncertainty restrict possible 2HDM
contributions (this procedure yields lower bounds on the charged-Higgs mass,
possibly also other constraints). An alternative approach would be to actually
fit the 2HDM to the data. This will not be discussed in the present paper.

This paper is organized as follows. In Sec.~\ref{sect:general} we discuss the
general constraints of positivity of the Higgs potential, together with
tree-level unitarity of Higgs--Higgs scattering amplitudes.  The impact of
these constraints is displayed in the $\tan\beta$--$M_{H^\pm}$ plane for a few
representative values of neutral Higgs boson masses.  Next, in
Sec.~\ref{sect:charged}, we discuss the constraints coming from the
$B$-physics experiments, in particular the $\bar B\to X_s\gamma$,
$B\to\tau\bar\nu_\tau$ and $B$--$\bar B$ oscillations. Sec.~\ref{sect:neutr}
is devoted to various experimental constraints that depend on details of the
neutral-Higgs sector. In Sec.~\ref{sect:combining}, we combine all the
constraints and in Sec.~\ref{sect:profile} we give some characteristics of the
surviving parameter space. In Sec.~\ref{sect:future} we speculate on possible
future experimental constraints, and then summarize in
Sec.~\ref{sect:summary}.

%%%%%%%%%%%%%%%%%%%%%%%%%%%%%%%%%%%%%%%%%%%%%%%%%%%%%%%%%%%%%%%%%%%%%%%%
\section{General theory constraints}
\setcounter{equation}{0}
\label{sect:general}
%%%%%%%%%%%%%%%%%%%%%%%%%%%%%%%%%%%%%%%%%%%%%%%%%%%%%%%%%%%%%%%%%%%%%%%%

In the general CP-non-conserving case, the neutral sector is conveniently
described by the three mixing angles, together with two masses $(M_1, M_2)$,
$\tan\beta$ and 
\begin{equation} \label{Eq:musq}
\mu^2=\frac{\Re m_{12}^2}{2\cos\beta\sin\beta}.
\end{equation}

We shall here project these constraints from the multi-dimensional parameter
space onto the $\tan\beta$--$M_{H^\pm}$ plane.  Such a projection of
information from a multi-dimensional space onto a point in the
$\tan\beta$--$M_{H^\pm}$ plane can be done in a variety of ways, all of which
will lead to some loss of information.  However, we feel that this loss of
detailed information can be compensated for by the ``overview'' obtained by
the following procedure:
\begin{enumerate}
\item
Pick a set of neutral-Higgs-boson masses, 
$(M_1, M_2)$ together with $\mu^2$.
\item
Scan an $N=n_1\times n_2\times n_3$ grid in
the $\alpha_1$--$\alpha_2$--$\alpha_3$ space, and count
the number $j$ of these points that give a viable model.
(Alternatively, one could scan over $N$ random points in this space.)
\item
The ratio
\begin{equation} \label{Eq:neutr_Q}
Q=j/N, \quad 0\le Q\le 1,
\end{equation}
is then a figure of merit, a measure of ``how allowed'' the point is, in the
$\tan\beta$--$M_{H^\pm}$ plane.  If $Q=0$, no sampled point in the
$\vecalpha=(\alpha_1,\alpha_2,\alpha_3)$ space is allowed, if $Q=1$, they are
all allowed.  An alternative measure
\begin{equation} \label{Eq:neutr_Q0}
Q_+=j/N_+,\quad Q_+\ge Q,
\end{equation}
counts in the denominator
only those points $N_+$ for which positivity is satisfied.
\end{enumerate}

Of course the 2HDM, if realized in nature, would only exist at {\it one} point
in this parameter space.  However, we think the above quantities $Q$ and $Q_+$
give meaningful measures of how ``likely'' different parameters are.

%%%%%%%%%%%%%%%%%%%%%%%%%%%%%%%%%%%%%%%%%%%%%%%%%%%%%%%%%%%%%%%%%%%%%%%%
\subsection{Reference masses}
\label{sect:refmasses}
%%%%%%%%%%%%%%%%%%%%%%%%%%%%%%%%%%%%%%%%%%%%%%%%%%%%%%%%%%%%%%%%%%%%%%%%
We shall impose the conditions of positivity, unitarity and experimental
constraints on the model, for the different ``reference'' mass sets given in
Table~\ref{tab:refmasses} (and variations around these).  For each of these
mass sets we scan the model properties in the
$\vecalpha=(\alpha_1,\alpha_2,\alpha_3)$ space.  (All scans have been
performed over a $200\times200\times100$ grid.  We note that other scanning
procedures might be more efficient \cite{Brein:2004kh}.) From these reference
masses, some trends will emerge, allowing us to draw more general conclusions.
%%%%%%%%%%%%%%%%%%%%%%%%%%%%%%%%%%%%%%%%%%%%%%%%%%%%%%%%%%%%%%%%%%%%%%
\begin{table}[ht]
\begin{center}
\renewcommand{\tabcolsep}{.75em}
\begin{tabular}{|c|c|c|c|c|}
\hline 
Name&$M_1 [\GeV]$&$M_2 [\GeV]$&$\mu^2~[\GeV]^2$\\
\hline
``100-300''&100&300&0 [$\pm(200)^2$]\\
``150-300''&150&300&0 [$\pm(200)^2$]\\
``100-500''&100&500&0 [$\pm(200)^2$]\\
``150-500''&150&500&0 [$\pm(200)^2$]\\
\hline
\end{tabular} 
\caption{Reference masses.}\label{tab:refmasses}
\end{center}
\end{table}
%%%%%%%%%%%%%%%%%%%%%%%%%%%%%%%%%%%%%%%%%%%%%%%%%%%%%%%%%%%%%%%%%%%%%%

In the $\text{2HDM}_5$, with $\Im\lambda_5\ne0$, the two input masses 
will together with $\vecalpha$
and $\tan\beta$ determine $M_3$ \cite{Khater:2003wq}. 
Specifying also $M_{H^\pm}$ and $\mu^2$,
all the $\lambda$'s can be determined.
(For explicit formulas, see \cite{Kaffas:2007rq}.)
%%%%%%%%%%%%%%%%%%%%%%%%%%%%%%%%%%%%%%%%%%%%%%%%%%%%%%%%%%%%%%%%%%%%%%%%
\subsection{Positivity and unitarity}
\label{sect:unitarity}
%%%%%%%%%%%%%%%%%%%%%%%%%%%%%%%%%%%%%%%%%%%%%%%%%%%%%%%%%%%%%%%%%%%%%%%%

Let us first discuss the effect of imposing positivity.  Actually, we will use
the term ``positivity'' to refer to the non-trivial conditions $M_3^2>0$ and
$M_2\le M_3$ together with $V(\Phi_1,\Phi_2)>0$ as $|\Phi_1|, |\Phi_2|
\to\infty$.  We shall henceforth refer to the set of points in the $\vecalpha$
space where positivity is satisfied, as $\vecalpha_+$.  In Table~II of
\cite{Kaffas:2007rq} we report percentages $Q$ of points in $\vecalpha$ space
for which positivity is satisfied. For the mass parameters of Table~I, the
fraction is around 30\%.  For small and negative values of $\mu^2$, ``most''
of the exclusion provided by the positivity constraint is due to the
conditions $M_3^2>0$ and $M_3>M_2$, without the explicit conditions on the
$\lambda$'s discussed, for example, in Appendix~A of \cite{ElKaffas:2006nt}.
%%%%%%%%%%%%%%%%%%%%%%%%%%%%%%%%%%%%%%%%%%%%%%%%%%%%%%%%%%%%%%%%%%%%%%
\begin{figure}[htb]
\begin{center}
\includegraphics[width=92mm]{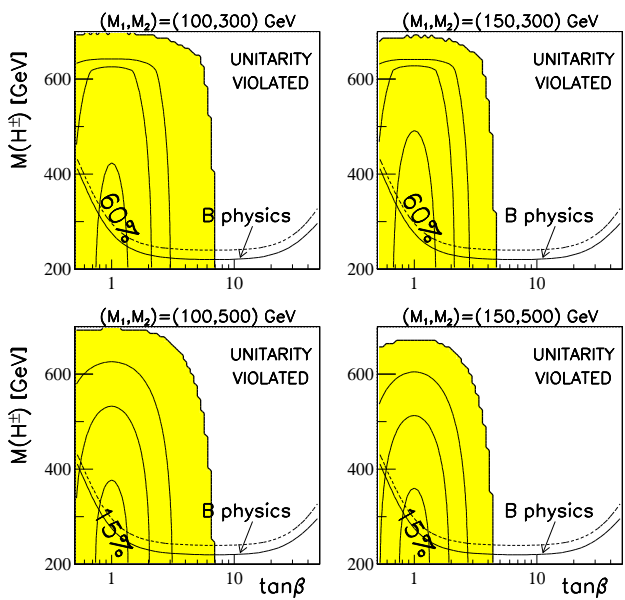}
\vspace*{-10mm}
\caption{\label{Fig:unitarity-mu=0} 
Percentage of points $Q_+$ in the $\vecalpha_+$ space that satisfy
unitarity.  
Four sets of $(M_1,M_2)$ values are considered, as indicated.
The contours show $Q_+=0$, 20\%, 40\%
and 60\% (upper panels) and 0, 5\%, 10\%  and 15\% (lower panels).
All panels: $\mu^2=0$. Yellow region: $Q_+>0$.
Also shown, are the 90\% and 95\% C.L.\ exclusion contours
from Fig.~\ref{Fig:chi2-general}.}
\end{center}
\vspace*{-5mm}
\end{figure}
%%%%%%%%%%%%%%%%%%%%%%%%%%%%%%%%%%%%%%%%%%%%%%%%%%%%%%%%%%%%%%%%%%%%%%

In Fig.~\ref{Fig:unitarity-mu=0}, for $\mu^2=0$, we study the effects of
imposing unitarity \cite{Kanemura:1993hm,Akeroyd:2000wc,Ginzburg:2003fe}.
This has a rather dramatic effect at ``large'' values of $\tan\beta$ and
$M_{H^\pm}$.  While the general constraints on the charged-Higgs sector, to be
discussed in Sec.~\ref{sect:charged}, exclude low values of $\tan\beta$ and
$M_{H^\pm}$, the constraints of unitarity exclude high values of these same
parameters.  Only some region in the middle remains not excluded.  For
$(M_1,M_2)=(100,300)$~GeV and $\mu^2=0$, unitarity excludes everything above
$\tan\beta\sim5$ (for any value of $M_{H^\pm}$), and above
$M_{H^\pm}\sim650$~GeV (for any value of $\tan\beta$).

We shall refer to the set of $\vecalpha$ values for which unitarity as well as
positivity are satisfied as $\hat\vecalpha\in\vecalpha_+$.  For $M_2=300$~GeV
(upper panels), the percentage of points in $\vecalpha_+$ space for which
unitarity is satisfied, reaches (at low $\tan\beta$ and low $M_{H^\pm}$)
beyond 60\%, whereas for $M_2=500$~GeV (lower panels), it only reaches values
of the order of 15--20\%.

The domains in which $Q_+>0$ depend on $\mu^2$: For negative values of
$\mu^2$, the region typically shrinks to lower values of $\tan\beta$, for
positive values of $\mu^2$ it extends to larger values of $\tan\beta$.  When
$\mu^2\simeq0$, and $\tan\beta\le5$, significant fractions of the
$\hat\vecalpha$ space are allowed.  However, for large positive values of
$\mu^2$, when also large values of $\tan\beta$ are allowed, only small domains
in $\hat\vecalpha$ remain allowed, as discussed in \cite{Kaffas:2007rq}. They
can actually be hard to find in a regular, equidistant scan over
$\hat\vecalpha$.

The unitarity constraints are conveniently formulated in terms of the
different weak isospin and hypercharge channels \cite{Ginzburg:2003fe}.  At
large values of $\tan\beta$ it turns out to be the isospin-zero,
hypercharge-zero channel that is most constraining.

%%%%%%%%%%%%%%%%%%%%%%%%%%%%%%%%%%%%%%%%%%%%%%%%%%%%%%%%%%%%%%%%%%%%%%%%
\section{General constraints from the $H_\pm$ sector}
\label{sect:charged}
\setcounter{equation}{0}
%%%%%%%%%%%%%%%%%%%%%%%%%%%%%%%%%%%%%%%%%%%%%%%%%%%%%%%%%%%%%%%%%%%%%%%%
The precise $B$-physics experiments provide severe constraints on the
charged-Higgs sector, excluding low values of $M_{H^\pm}$ and $\tan\beta$.
We shall here discuss the three most severe constraints of this kind
that are independent of the neutral sector.
%%%%%%%%%%%%%%%%%%%%%%%%%%%%%%%%%%%%%%%%%%%%%%%%%%%%%%%%%%%%%%%%%%%%%%%%
\subsection{$\bar B\to X_s\gamma$}
%%%%%%%%%%%%%%%%%%%%%%%%%%%%%%%%%%%%%%%%%%%%%%%%%%%%%%%%%%%%%%%%%%%%%%%%

The $\bar B\to X_s\gamma$ branching rate was early found to 
constrain the allowed charged-Higgs-boson masses, but also to be very sensitive
to QCD effects. At leading logarithmic order (LO), it is given by
\cite{Bertolini:1986th,Grinstein:1987pu,Grigjanis:1988iq}:
\begin{align} \label{Eq:B-to-sgamma-0}
{\cal B}(\bar B\to X_s\gamma)
&=\frac{|V_{ts}^\ast V_{tb}|^2}{|V_{cb}|^2}\,
\frac{6\alpha_\text{e.m.}}{\pi g(z)}|C_7^{(0)\text{eff}}(\mu_b)|^2 \nonumber \\
&\times{\cal B}(\bar B\to X_c e\bar\nu_e),
\end{align}
where the first factor is a ratio of CKM matrix elements, $g(z=m_c^2/m_b^2)$
is a phase space factor, and $C_7^{(0)\text{eff}}(\mu_b)$ is an effective
Wilson coefficient, evaluated at the $B$-meson scale, $\mu_b$.  This effective
Wilson coefficient is obtained from the relevant ones at the electroweak
scale, $C_i^{(0)}(\mu_0)$, $i=2,7,8$ \cite{Buras:1993xp}, where the effects of
the 2HDM enter. Certain linear combinations are denoted ``effective''
coefficients, they are defined such that e.g.\ $C_7^{(0)\text{eff}}(\mu)$
includes all one-loop contributions to $b\to s\gamma$, i.e., also those of
four-quark operators \cite{Buras:1993xp}.  The effective Wilson coefficient
was early found to be quite sensitive to the scale relevant to $B$-meson
decay, changing by $\pm25\%$ if the scale $\mu_b$ is varied by a factor of two
in either direction around $m_b\simeq5$~GeV \cite{Ali:1993ct,Buras:1993xp}.
At the NLO, this scale sensitivity is however significantly reduced
\cite{Adel:1993ah,Misiak:1992bc,Greub:1996jd,Chetyrkin:1996vx,
Buras:1997bk,Ciuchini:1997xe,Gambino:2001ew,Buras:2002tp,Asatrian:2005pm}.

The additional contributions due to the 2HDM can at the weak scale be
described by diagrams involving $H^\pm$ exchange, and depend on this mass,
$M_{H^\pm}$, as well as on the Yukawa couplings, i.e., on $\tan\beta$.  At
leading logarithmic order in QCD, they are discussed in
\cite{Grinstein:1987pu,Barger:1989fj}.  Some of these additional terms can be
enhanced by factors $\cot^2\beta$ from the $H^\pm$ Yukawa coupling squared.
They all vanish linearly in $m_t^2/M_{H^\pm}^2$, i.e., for $M_{H^\pm}^2\gg
m_t^2$.

The NLO results for the 2HDM have been studied by many authors, see
\cite{Ciuchini:1994xa,Ciuchini:1997xe,Borzumati:1998tg,Gambino:2001ew}.  As
compared with the LO calculation, it has been found that the NLO effects
weaken the constraints on the allowed region in the $\tan\beta$--$M_{H^\pm}$
plane \cite{Buras:1993xp,Ciuchini:1994xa,Borzumati:1998tg}, the bound on
$M_{H^\pm}$ is significantly relaxed.

Apart from minor effects, this calculation has for the SM been carried to the
next-to-next-to-leading order (NNLO) \cite{Misiak:2006zs}.  
At the NNLO, the branching ratio can be written as
\cite{Gambino:2001ew,Misiak:2006ab}:
\begin{align} \label{Eq:B-to-sgamma}
{\cal B}(\bar B\to X_s\gamma)
&=\frac{|V_{ts}^\ast V_{tb}|^2}{|V_{cb}|^2}\,
\frac{6\alpha_\text{e.m.}}{\pi C}\bigl\{P(E_0)+N(E_0)\bigr\} \nonumber \\
&\quad\times
{\cal B}(\bar B\to X_c e\bar\nu_e)_\text{exp},
\end{align}
where $P$ and $N$ denote perturbative and non-per\-turbative effects that both
depend on the photon lower cut-off energy $E_0$.  In the LO limit, the term
$P$ reduces to the square of the effective Wilson coefficient
$C_7^{(0)\text{eff}}(\mu_b)$ in (\ref{Eq:B-to-sgamma-0}), whereas NLO and NNLO
contributions include effects due to gluon exchange and emission, and require
the summation over a bilinear expression involving also other Wilson
coefficients. The factor $C$ accounts for $m_c$-dependence associated with the
semileptonic decay $\bar B\to X_c e\bar\nu_e$ \cite{Gambino:2001ew}.

The SM prediction of Misiak {\it et al.}  is $(3.15\pm0.23)\times10^{-4}$ for
$E_0=1.6~\text{GeV}$ \cite{Misiak:2006zs,Misiak:2006ab}, if all errors are
added in quadrature. This is to be compared with the recent experimental
results, which are averaged to $3.55\times10^{-4}$ \cite{Barberio:2006bi},
with an uncertainty of $7$--$7.5\%$, again with statistical and systematic
errors added in quadrature.  Andersen and Gardi \cite{Andersen:2006hr}
advocate a different approach to the resummation of the perturbation series,
by ``Dressed Gluon Exponentiation'', which accounts for multiple and soft
collinear radiation.  These effects are particularly important for large
photon energies.  Their approach yields $(3.47\pm0.48)\times10^{-4}$, i.e.,
they find a 14\% uncertainty.  Becher and Neubert also introduce further
corrections to the calculation of the photon spectrum, and find a rather low
value, $(2.98\pm0.26)\times10^{-4}$ \cite{Becher:2006pu}, leaving more room
for new physics.

The 2HDM contribution is positive, a finite value for the charged Higgs mass
would thus bring the results of \cite{Misiak:2006zs,Becher:2006pu} in closer
agreement with the experiment. We shall however take the attitude that these
numbers are compatible and compare the uncertainty in the experimental result
and the SM prediction with the 2HDM contribution.

In the NNLO, the perturbative contribution $P$ is obtained via
the following three steps \cite{Misiak:2006ab}:
\begin{enumerate}
\item
Evaluation of the Wilson coefficients at the ``high'' (electroweak) scale,
$\mu_0$ \cite{Bobeth:1999mk,Misiak:2004ew}.  These coefficients are expanded
to second order in $\alpha_s$ and rotated to ``effective'' Wilson coefficients
$C_i^\text{eff}(\mu_0)$ \cite{Buras:1993xp,Czakon:2006ss}.  The 2HDM effects
enter at this stage, at lowest order in the Wilson coefficients $C_7(\mu_0)$
and $C_8(\mu_0)$.
\item
Evaluation of the ``running'' and mixing of these operators, from the high
scale to the ``low'' ($B$-meson) scale.  This is where the main QCD effects
enter via a matrix $U$ that is given in terms of powers of
$\eta=\alpha_s(\mu_0)/\alpha_s(\mu_b)$ \cite{Buras:1993xp,
Chetyrkin:1996vx,Gorbahn:2004my,Czakon:2006ss}.
\item
Evaluation of matrix elements at the low scale
\cite{Blokland:2005uk,Misiak:2006ab}, which amounts to constructing $P(E_0)$
of Eq.~(\ref{Eq:B-to-sgamma}) from the $C_i^\text{eff}(\mu_b)$.
\end{enumerate}

We adopt the scale parameters of \cite{Misiak:2006zs}:
\begin{equation} \label{Eq:scales}
\mu_0=160~\text{GeV}, \quad
\mu_b=2.5~\text{GeV}, \quad
\mu_c=1.5~\text{GeV}.
\end{equation}

Actually, our treatment of the higher-order effects has been simplified
compared to that described in \cite{Misiak:2006zs,Misiak:2006ab}, in the sense
that: $(i)$ We determine the contribution $P_2^{(2)\beta_0}$ of
\cite{Misiak:2006ab} by their Eq.~(4.10), using results of
\cite{Buras:2002tp,Blokland:2005uk,vanRitbergen:1999gs}, but read off
$P_2^{(2)\text{rem}}=5$ (valid at the ``default'' scales) from their Fig.~2.
$(ii)$ Rather than explicitly including the ${\cal O}(V_{ub})$ and electroweak
corrections, we adopt the corresponding numerical values of +1\%
\cite{Gambino:2001ew,Misiak:2006ab} and $-3.7$\%
\cite{Gambino:2001ew,Gambino:2001au}, respectively.

Furthermore, while the ``matching'' in the SM is performed to second order in
$\alpha_s$, we include, in addition to the dominant, lowest order, 2HDM
effects, also the first-order (in $\alpha_s(\mu_0)$) contributions.  In fact,
we include the latter at the level of ``effective'' Wilson coefficients,
following \cite{Borzumati:1998tg}, and take their ``matching scale'' $\mu_W$
as $\mu_0$.  Characteristic relative magnitudes of the LO and NLO 2HDM
contributions are given in Table~\ref{tab:BR-2HDM}, for $\tan\beta=1$ and 10,
and $M_{H^\pm}=300~\text{GeV}$ and 600~GeV. The LO 2HDM contribution is
measured with respect to the full SM value, whereas the NLO 2HDM contribution
(according to \cite{Borzumati:1998tg}) is measured relative to the SM value
plus the LO 2HDM contribution. The missing ${\cal O}(\alpha_s^2(\mu_0))$
corrections appear unimportant.

%%%%%%%%%%%%%%%%%%%%%%%%%%%%%%%%%%%%%%%%%%%%%%%%%%%%%%%%%%%%%%%%%%%%%%
\begin{table}[ht]
\begin{center}
\renewcommand{\tabcolsep}{.75em}
\begin{tabular}{|c|c|c|c|c|c|}
\hline 
$M_{H^\pm}$&\multicolumn{2}{c|}{$\tan\beta=1$}
&\multicolumn{2}{c|}{$\tan\beta=10$} \\
\cline{2-5}
[GeV] & LO & NLO & LO & NLO \\
\hline
600&18.2&$-3.7$&16.3&$-3.5$\\
300&41.8&$-5.5$&36.5&$-5.1$\\
\hline
\end{tabular} 
\caption{Relative (in per cent) 2HDM contributions to the
branching ratio at LO and NLO (Borzumati \& Greub).}
\label{tab:BR-2HDM}
\end{center}
\end{table}
%%%%%%%%%%%%%%%%%%%%%%%%%%%%%%%%%%%%%%%%%%%%%%%%%%%%%%%%%%%%%%%%%%%%%%

As input parameters for (\ref{Eq:B-to-sgamma}), in addition to
(\ref{Eq:scales}), we take the CKM ratio to be 0.9676 \cite{Charles:2004jd},
${\cal B}(\bar B\to X_c e\bar\nu_e)=0.1061$ \cite{Aubert:2004aw} (see also
\cite{Koppenburg:2004fz}), $m_t(\mu_0)=162$~GeV (corresponding to a pole mass
of $m_t=171.4$~GeV), $m_c(m_c)=1.224$~GeV \cite{Huber:2005ig} and
$m_b^{1S}=4.68$~GeV \cite{Huber:2005ig}, yielding
$m_c(\mu_c)=1.131~\text{GeV}$ and $m_c(\mu_c)/m_b=0.242$.  For the
non-perturbative part, $N(E_0)$, we follow refs.\ \cite{Misiak:2006ab} and
\cite{Bauer:1997fe,Neubert:2004dd}.  Our result for the branching ratio,
Eq.~(\ref{Eq:B-to-sgamma}), in the SM limit ($M_{H^\pm}\to\infty$), is
$3.12\times10^{-4}$.  We collect in Table~\ref{tab:bsgamma} the various values
obtained by different authors.

%%%%%%%%%%%%%%%%%%%%%%%%%%%%%%%%%%%%%%%%%%%%%%%%%%%%%%%%%%%%%%%%%%%%%%
\begin{table}[ht]
\begin{center}
\renewcommand{\tabcolsep}{.75em}
\begin{tabular}{|c|c|c|c|c|}
\hline 
Authors&${\cal B}$&$\sigma$&reference\\
\hline
HFAG &3.55&$0.24\pm\cdots$&\cite{Barberio:2006bi}\\
M.M. {\it et al.}&3.15&0.23&\cite{Misiak:2006zs}\\
J. A. \& E. G.&3.47&0.48&\cite{Andersen:2006hr}\\
T. B. \& M. N.&2.98&0.26&\cite{Becher:2006pu}\\
present work&3.12&--& \\
\hline
\end{tabular} 
\caption{Branching ratios ${\cal B}(\bar B\to X_s\gamma)$ 
and uncertainties in units of $10^{-4}$, all for
$E_\gamma>E_0=1.6~\text{GeV}$.}
\label{tab:bsgamma}
\end{center}
\end{table}
%%%%%%%%%%%%%%%%%%%%%%%%%%%%%%%%%%%%%%%%%%%%%%%%%%%%%%%%%%%%%%%%%%%%%%

We define a $\chi^2$ measure of the amount by which the 2HDM would violate the
agreement with the SM value, 
\begin{equation} \label{Eq:chi2_bsgamma}
\chi^2_{b\to s\gamma}=\frac{[{\cal B}(\bar B\to X_s\gamma)_\text{2HDM}
-{\cal B}(\bar B\to X_s\gamma)_\text{ref}]^2}
{\{\sigma[{\cal B}(\bar B\to X_s\gamma)]\}^2}.
\end{equation}
Here, ${\cal B}(\bar B\to X_s\gamma)_\text{2HDM}$ denotes the 2HDM prediction.
As noted above, it will depend on $\tan\beta$ and $M_{H^\pm}$, whereas ${\cal
B}(\bar B\to X_s\gamma)_\text{ref}$ denotes a reference value, taken to be the
averaged experimental value, $3.55\times10^{-4}$ \cite{Barberio:2006bi}.

For the uncertainty that enters in (\ref{Eq:chi2_bsgamma}), we adopt the value 
\begin{equation} \label{Eq:bsgamma-sigma}
\sigma[{\cal B}(\bar B\to X_s\gamma)]=0.35\times10^{-4},
\end{equation}
which corresponds to the experimental and (SM) theoretical uncertainties
\cite{Misiak:2006zs} added in quadrature.  For the 2HDM, the theoretical
studies have not been carried to the same level of precision, but we recall
that the 2HDM-specific NLO-level contributions only modify the over-all
branching ratios by ${\cal O}(5\%)$.  These numbers give
$\chi^2_\text{min}=1.52$.

\begin{figure}[htb]
\vspace*{-5mm}
\begin{center}
\includegraphics[width=7cm]{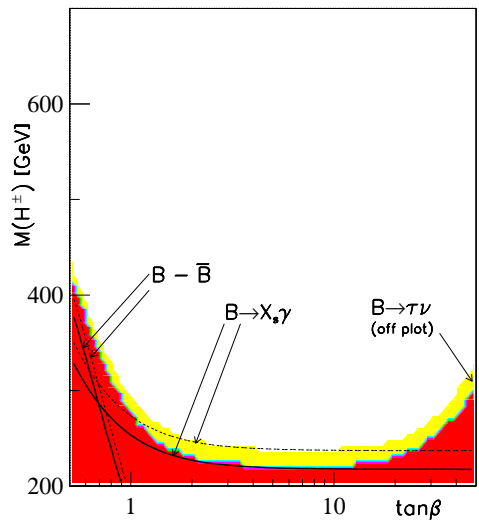}
\vspace*{-4mm}
\caption{\label{Fig:chi2-general} Excluded regions at small $M_{H^\pm}$ and
small $\tan\beta$ due to constraints from $\bar B\to X_s\gamma$ and $B$--$\bar
B$ oscillations.  Dashed: 90\% C.L., solid: 95\% C.L.  The corresponding
curves for $B\to\tau\nu$ are off the plot, to the lower right.  Colored: joint
exclusion at 90\% and 95\% C.L.}
\end{center}
\vspace*{-4mm}
\end{figure}

With the above choice, we can determine exclusion regions in the
$\tan\beta$--$M_{H^\pm}$ plane.  Such regions are shown in
Fig.~\ref{Fig:chi2-general} for $\chi^2=4.61$ and 5.99, corresponding to
confidence levels of 90\% and 95\%, respectively, for 2 degrees of freedom
($\tan\beta$ and $M_{H^\pm}$).
The obtained bound differs from that given by Misiak {\it et al}.\
\cite{Misiak:2006zs} even though the branching ratio is reproduced to a
reasonable precision. The reason is as follows. For compatibility with the
treatment of the other experimental constraints, we consider a {\it
two-dimensional} constraint, whereas they consider a single-sided
one-dimensional constraint.\footnote{We shall refer to $\chi^2=4.61$
and 5.99 as 90\% and 95\% C.L., respectively, even though this conversion
to a probability only applies to the simple, ideal, case.}

%%%%%%%%%%%%%%%%%%%%%%%%%%%%%%%%%%%%%%%%%%%%%%%%%%%%%%%%%%%%%%%%%%%%%%
%%%%%%%%%%%%%%%%%%%%%%%%%%%%%%%%%%%%%%%%%%%%%%%%%%%%%%%%%%%%%%%%%%%%%%

By adopting the approximate SM-value $P_2^{(2)\text{rem}}=5$,
we only include part of the 2HDM-specific contributions to $P^{(2)}$.
A more well-defined procedure would be to leave out {\it all}
2HDM-specific contributions to $P^{(2)}$.\footnote{We are grateful
to M. Misiak for discussions on this and related issues.} However,
that procedure leads to an exclusion limit about 30~GeV higher
in $M_{H^\pm}$, i.e., more of the parameter space would be excluded.
We have chosen the more conservative approach, which leads to less
exclusion.
%%%%%%%%%%%%%%%%%%%%%%%%%%%%%%%%%%%%%%%%%%%%%%%%%%%%%%%%%%%%%%%%%%%%%%%%
\subsection{$B\to X\tau\bar\nu_\tau$ and $B^-\to\tau\bar\nu_\tau$}
\label{Sec:B-tau-nu}
%%%%%%%%%%%%%%%%%%%%%%%%%%%%%%%%%%%%%%%%%%%%%%%%%%%%%%%%%%%%%%%%%%%%%%%%
Charged Higgs bosons would contribute to the decay
\begin{equation} \label{Eq:btoctau}
b\to c\tau\bar\nu_\tau
\end{equation}
from which the bound
\begin{equation} \label{Eq:btoctau-bound}
\tan\beta<0.52~\GeV\times M_{H^\pm}
\end{equation}
has been obtained \cite{Hou:1992sy} at the $2\sigma$ level.
This bound, which would exclude a corner to the lower right
outside the region shown in Fig.~\ref{Fig:chi2-general} is actually 
irrelevant, since such high values of $\tan\beta$ are excluded by
the other constraints to be discussed in Sec.~\ref{sect:neutr}.

Charged Higgs-boson exchange also contributes to
\begin{equation}
B^-\to\tau\bar\nu_\tau
\end{equation}
which similarly provides a bound of the kind (\ref{Eq:btoctau-bound}).  The
measurement gives ${\cal B}(B^-\to\tau\bar\nu_\tau=(1.79\pm0.71)\times10^{-4}$
\cite{Ikado:2006un}, where we have added in quadrature symmetrized statistical
and systematic errors. With a Standard-Model prediction of 
$(1.59\pm0.40)\times10^{-4}$,
\begin{equation}
r_{H\,\text{exp}}=\frac{{\cal B}(B^-\to\tau\bar\nu_\tau)}
{{\cal B}(B^-\to\tau\bar\nu_\tau)_\text{SM}}
=1.13\pm0.53.
\end{equation}
Interpreted in the framework of the 2HDM, one finds \cite{Hou:1992sy}
\begin{equation}
r_{H\,\text{2HDM}}=\biggl[1-\frac{m_B^2}{M_{H^\pm}^2}\,\tan^2\beta\biggr]^2.
\end{equation}
We take $m_B=5.28~\GeV$, and formulate a $\chi^2$ measure as
\begin{equation}
\chi^2_{b\to\tau\nu}=\frac{[r_{H\,\text{2HDM}}-r_{H\,\text{exp}}]^2}
{[\sigma(r_{H\,\text{exp}})]^2}.
\end{equation}
It follows that two sectors at large values of $\tan\beta$ and low
values of $M_{H^\pm}$ will be excluded.  This bound is stronger than that of
(\ref{Eq:btoctau-bound}).  It has some relevance in the
lower right of Fig.~\ref{Fig:chi2-general}, when all effects are added
(see Sec.~\ref{sect:charged-combined}).
%%%%%%%%%%%%%%%%%%%%%%%%%%%%%%%%%%%%%%%%%%%%%%%%%%%%%%%%%%%%%%%%%%%%%%%%
\subsection{$B$--$\bar B$ oscillations}
%%%%%%%%%%%%%%%%%%%%%%%%%%%%%%%%%%%%%%%%%%%%%%%%%%%%%%%%%%%%%%%%%%%%%%%%

The precisely measured $B_d$--$\bar B_d$ oscillations
are at lowest order given by the formula \cite{Abbott:1979dt}
\begin{equation} \label{Eq:Delta_m_B_d}
\Delta m_{B_d}
=\frac{G_\text{F}^2}{6\pi^2}
|V_{td}^\ast|^2|V_{tb}|^2f_B^2\,B_B\, m_B\,\eta\,M_W^2\,S_{2HDM},
\end{equation}
with the Inami--Lim functions \cite{Inami:1980fz}
\begin{equation} \label{Eq:Inami--Lim}
S_{2HDM}=S_{WW}+2S_{WH}+S_{HH}.
\end{equation}
The contributions proportional to diagrams with the exchange
of one or two charged-Higgs-bosons are denoted
$S_{WH}$ and $S_{HH}$. 
They are proportional to $\cot^2\beta$ and $\cot^4\beta$, 
and vanish when $M_{H^\pm}\gg m_t$, as 
$(m_t^2/M_{H^\pm}^2)\log(m_t^2/M_{H^\pm}^2)$ and $m_t^2/M_{H^\pm}^2$,
respectively.

At the NLO, the corresponding result has also been obtained
\cite{Urban:1997gw}: $B_B$ and $\eta$ (denoted $\eta_2$ in
\cite{Urban:1997gw}) receive corrections of order $\alpha_s$. In particular,
the ${\cal O}(\alpha_s)$ contribution to $\eta_2$ becomes a non-trivial
function of $\tan\beta$ and $M_{H^\pm}$.  A few comments are here in order:
(i) The factor of 2 for the second term in (\ref{Eq:Inami--Lim}) has been
adopted to follow the notation of \cite{Urban:1997gw}. (ii) There is a
book-keeping problem with the expression for $L^{(i,H)}$ in Eq.~(A.20) of
\cite{Urban:1997gw}. Since the last term in that expression, proportional to a
quantity denoted $HH$, has an explicit coefficient $1/\tan^4\beta$, the
$S_{HH}$ in Eqs.~(A.21) and (A.24) for $HH^{(i)}$ should be replaced by
$\tan^4\beta\times S_{HH}$. (iii) There is a discrepancy in a quantity denoted
$2WW_{tu}^{(8)}$, between Eq.~(A.16) in \cite{Urban:1997gw} and the later PhD
thesis of the same author. We have chosen to take the formula given in the
thesis.  At the level of $\eta_2$, it amounts to a difference of the order of
2\%.\footnote{We are grateful to A. Buras, U. Jentschura and F. Krauss for
correspondence on these points (ii) and (iii).}

The ${\cal O}(\alpha_s)$ corrections to $\eta$ introduce a variation of $\eta$
(or $\eta_2$) from 0.334 at $\tan\beta=0.5$ and $M_{H^\pm}=200~\GeV$ to 0.552
at $\tan\beta=50$. (We have adopted the over-all normalization
of $\eta$ such that it agrees with the SM value 0.552 \cite{Ball:2006xx}
at $\tan\beta=50$.) However, the product
$\eta\times S_{2HDM}$ varies by a factor of 2.7 over this same range,
as compared with a factor of 4.5 for $S_{2HDM}$ itself. Thus, the
inclusion of the ${\cal O}(\alpha_s)$ QCD corrections reduce the sensitivity
of $\Delta m_B$ to charged-Higgs contributions.
In other words, the inclusion of NLO corrections weakens the constraints
on the 2HDM at low values of $\tan\beta$.

Recently, also the $B_s$--$\bar B_s$ oscillation parameter $\Delta m_{B_s}$
has been measured \cite{Abulencia:2006mq}.
It is given by an expression similar to (\ref{Eq:Delta_m_B_d}), except that
the CKM matrix elements are different:
\begin{equation} \label{Eq:Delta_m_B_s}
\Delta m_{B_s}
=\frac{G_\text{F}^2}{6\pi^2}
|V_{ts}^\ast|^2|V_{tb}|^2f_B^2\,B_B\, m_B\,\eta\,M_W^2\,S_{2HDM},
\end{equation}
and $f_B^2\,B_B$ and $m_B$, now referring to $B_s$, are numerically different.

The quantities $f_B\,B_B^{1/2}$ that appear in (\ref{Eq:Delta_m_B_d})
and (\ref{Eq:Delta_m_B_s}) are determined from lattice QCD studies,
and rather uncertain. We shall use the values adopted recently by Ball
and Fleischer \cite{Ball:2006xx},
\begin{align} \label{Eq:f_B_d_etc}
f_{B_d}\,B_{B_d}^{1/2}
&=(0.244\pm0.026)~\GeV, \nonumber \\
f_{B_s}\,B_{B_s}^{1/2}
&=(0.295\pm0.036)~\GeV,
\end{align}
based on unquenched calculations.
In fact, these uncertainties are the dominant ones.

For each of these observables, one can form a $\chi^2$ measure of the deviation
from the SM:
\begin{equation}
\chi^2_{B_q-\bar B_q}
=\frac{(\Delta m_{B{_q}\text{2HDM}}-\Delta m_{B{_q}\text{SM}})^2}
{[\sigma(\Delta m_{B_q})]^2}.
\end{equation}
We adopt the attitude that the measurements of $\Delta m_{B_d}$ and $\Delta
m_{B_s}$ furnish determinations of the CKM matrix elements $|V_{td}|$ and
$|V_{ts}|$ that are compatible with the SM, and simply require that the
additional 2HDM contributions do not spoil this consistency. With the
assumption that the uncertainties in (\ref{Eq:f_B_d_etc}) are the dominant
ones, we have
\begin{align} \label{Eq:Delta_m_B_errors}
\sigma(\Delta m_{B_d})=21\%\times \Delta m_{B_d}, \nonumber \\
\sigma(\Delta m_{B_s})=24\%\times \Delta m_{B_s}.
\end{align}
However, with the theory error (\ref{Eq:f_B_d_etc}) being the dominant one, we
can not claim that the measurements of $\Delta m_{B_d}$ and $\Delta m_{B_s}$
furnish independent constraints on the model. Therefore, we consider
only the $\chi^2$ contribution from the more constraining one of these two
measurements, namely $\Delta m_{B_d}$ with a 21\% uncertainty:
\begin{equation}
\chi^2_{B-\bar B}=\chi^2_{B_d-\bar B_d}.
\end{equation}
Contours of $\chi^2_{B-\bar B}$ are shown in Fig.~\ref{Fig:chi2-general}.
As mentioned above, they are more generous than the corresponding
bounds based on the LO theory.
%%%%%%%%%%%%%%%%%%%%%%%%%%%%%%%%%%%%%%%%%%%%%%%%%%%%%%%%%%%%%%%%%%%%%%%%
\subsection{Combining general constraints from the $H^\pm$ sector}
\label{sect:charged-combined}
%%%%%%%%%%%%%%%%%%%%%%%%%%%%%%%%%%%%%%%%%%%%%%%%%%%%%%%%%%%%%%%%%%%%%%%%

In order to quantify the extent to which a particular point in the
$\tan\beta$--$M_{H^\pm}$ plane is forbidden, we form a $\chi^2$ as follows:
\begin{equation} \label{Eq:chi2-general}
\chi^2_\text{general}
=\chi^2_{b\to s\gamma}+\chi^2_{b\to\tau\nu}+\chi^2_{B-\bar B}.
\end{equation}
The subscript ``general'' refers to the fact that constraints depending on the
neutral sector are not yet taken into account. Yet, no choice of
neutral-sector parameters can avoid these constraints.

Figure \ref{Fig:chi2-general} shows excluded regions in the
$\tan\beta$--$M_{H^\pm}$ plane, due to the constraints of $\bar B\to
X_s\gamma$, $B^-\to\tau\bar\nu_\tau$ and $B$--$\bar B$ oscillations at 90\%
(dashed) and 95\% C.L.\ (solid), as well as combined (indicated in yellow and
red at 90\% and 95\% C.L., respectively).  Earlier versions of such exclusion
plots can be found in \cite{Grant:1994ak,Gambino:2001ew,Cheung:2003pw}.

%%%%%%%%%%%%%%%%%%%%%%%%%%%%%%%%%%%%%%%%%%%%%%%%%%%%%%%%%%%%%%%%%%%%%%%%
\section{Experimental constraints on the neutral sector} 
\setcounter{equation}{0}
\label{sect:neutr}
%%%%%%%%%%%%%%%%%%%%%%%%%%%%%%%%%%%%%%%%%%%%%%%%%%%%%%%%%%%%%%%%%%%%%%%%

We now turn to the constraints coming from experiments related to the neutral
Higgs sector. These are not ``general'', they will depend on the parameters of
this sector, namely neutral Higgs masses and mixing angles. We will study, as
representative cases, those given in Table~\ref{tab:refmasses}.

A problem with point~2 of the procedure of Sec.~\ref{sect:general}, is the
following.  The conditions of positivity and unitarity are
absolute\footnote{One might consider models for which perturbative unitarity
is not satisfied, but we shall not do that here.}, with a certain fraction
$Q_+$ of the points in the $\vecalpha_+$ space satisfying these, whereas the
experimental constraints are statistical, i.e., a certain parameter point may
violate an experimental observation by 1$\sigma$, $2\sigma$ or more.  There is
no obvious way to combine this information.

As a first step, we will consider one of these experimental constraints on the
neutral sector at a time, and display the way it excludes parts of the
$\tan\beta$--$M_{H^\pm}$ plane by the following simple consideration. Let
$\chi_i^2$ be the contribution to $\chi^2$ due to a particular experimental
observable ${\cal O}_i$:
\begin{equation} \label{Eq:chi2-ind-neutr}
\chi^2_i
=\frac{({\cal O}_{i,\text{2HDM}}-{\cal O}_{i,\text{ref}})^2}
{[\sigma({\cal O}_{i})]^2},
\end{equation}
where the ``reference'' value ${\cal O}_{i,\text{ref}}$ will be either the
experimental value or the SM value.

We shall consider the following observables: $R_b$, the branching ratio for
$Z\to b\bar b$ \cite{Yao:2006px} (in Sec.~\ref{sect:neutr-Gamma_b}); the LEP2
non-discovery of a light neutral Higgs boson
\cite{Boonekamp:2004ae,Achard:2003ty} (in
Sec.~\ref{sect:neutr-LEP-non-discovery}); $\rho$, the LEP determination of the
relation between the $Z$ and $W$ masses \cite{Ross:1975fq} (in
Sec.~\ref{sect:neutr-Deltarho}); and $a_\mu=\half(g-2)_\mu$, the precise
Brookhaven determination of the muon anomalous magnetic moment
\cite{Bennett:2004pv} (in Sec.~\ref{sect:neutr-g-2}).  Then, the criterion we
will consider in this first step, can be expressed in terms of the following
quantity:
\begin{itemize}
\item
For fixed $\tan\beta$ and $M_{H^\pm}$ take
\begin{equation}
\hat\chi_i^2=\mathop{\min}_{\hat\vecalpha\in\vecalpha_+}\chi_i^2
\end{equation}
where $\chi_i^2$ is minimized over the part $\hat\vecalpha$ of the
$\vecalpha_+$ space for which positivity and also unitarity are satisfied.
\end{itemize}
The minimization over $\hat\vecalpha$ finds the point where the model can most
easily accommodate the particular experimental constraint, with the chosen
masses ($M_1,M_2$) and $\mu^2$ fixed.  This point in $\hat\vecalpha$ will in
general differ from one experimental constraint to another.  We will establish
90 and 95\% C.L.\ allowed regions in the $\tan\beta$--$M_{H^\pm}$ plane,
corresponding to one of these observables at a time. These will subsequently
be combined. However, the over-all allowed regions will in general be less
than the intersection of the individual ones, since the latter will correspond
to different points in $\hat\vecalpha$.

This procedure preserves none of the probabilistic information of
Fig.~\ref{Fig:unitarity-mu=0}, which shows that some region of the
$\tan\beta$--$M_{H^\pm}$ plane contains more possible solutions than some
other region. All focus is here on the one ``best'' point in $\vecalpha$,
where $\chi^2$ is lowest.

%%%%%%%%%%%%%%%%%%%%%%%%%%%%%%%%%%%%%%%%%%%%%%%%%%%%%%%%%%%%%%%%%%%%%%%%
\subsection{$R_b$ constraint (LEP)} 
\label{sect:neutr-Gamma_b}
%%%%%%%%%%%%%%%%%%%%%%%%%%%%%%%%%%%%%%%%%%%%%%%%%%%%%%%%%%%%%%%%%%%%%%%%

The 2HDM-specific contributions to $R_b$ are of two kinds.  At low values of
$\tan\beta$, the exchange of charged Higgs bosons is important, whereas at
high values of $\tan\beta$ the exchange of neutral Higgs bosons is important
\cite{Denner:1991ie}.  For the general CP non-conserving case, the latter
contributions are given in \cite{ElKaffas:2006nt}.

%%%%%%%%%%%%%%%%%%%%%%%%%%%%%%%%%%%%%%%%%%%%%%%%%%%%%%%%%%%%%%%%%%%%%%
\begin{figure}[htb]
\vspace*{-2mm}
\begin{center}
\includegraphics[width=95mm]{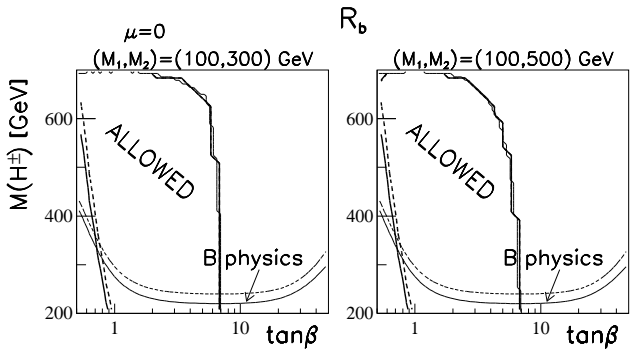}
\vspace*{-10mm}
\caption{\label{Fig:delga-mu=0} Exclusions due to the $\Delta R_b$
constraint, $\hat\chi_{R_b}^2$, for $(M_1,M_2)=(100,300)$~GeV, $\mu=0$.
Also shown is the region excluded by the $B$-physics constraints, and the 0\%
contour from Fig.~\ref{Fig:unitarity-mu=0}.}
\end{center}
\end{figure}
%%%%%%%%%%%%%%%%%%%%%%%%%%%%%%%%%%%%%%%%%%%%%%%%%%%%%%%%%%%%%%%%%%%%%%

For this observable, the reference value in (\ref{Eq:chi2-ind-neutr}) is not
well defined, since $R_b$ \cite{:2005em} is part of the electroweak
observables from which a SM Higgs mass is fitted.  Hence, we take
\begin{equation}
\chi^2_{R_b}
= \biggl(\frac{\Delta R_{b,}{}_\text{2HDM}}
{\sigma(R_b)}\biggr)^2,
\end{equation}
where $\Delta R_{b,}{}_\text{2HDM}$ refers to the 2HDM-specific contributions
to this quantity \cite{Denner:1991ie,ElKaffas:2006nt} and with
$\sigma(R_b)=0.05\%$ the experimental uncertainty \cite{Yao:2006px}.

We show in Fig.~\ref{Fig:delga-mu=0}, for two sets of $(M_1,M_2)$ values, and
$\mu=0$, how this constraint removes a sliver of low-$\tan\beta$ values.
The allowed regions are here cut off at $\tan\beta\gsim7$ due to the unitarity
constraint discussed in Sec.~\ref{sect:unitarity}. The neutral-Higgs-exchange
contribution to $R_b$ is in this case of no importance.

%%%%%%%%%%%%%%%%%%%%%%%%%%%%%%%%%%%%%%%%%%%%%%%%%%%%%%%%%%%%%%%%%%%%%%
\begin{figure}[htb]
\vspace*{-2mm}
\begin{center}
\includegraphics[width=95mm]{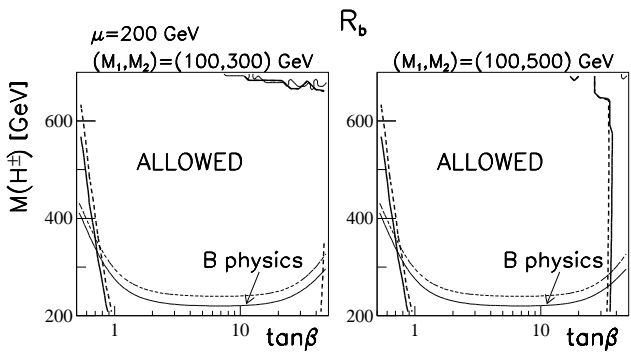}
\vspace*{-10mm}
\caption{\label{Fig:delga-mu=200}
$\Delta R_b$ constraint, $\hat\chi_{R_b}^2$, for
$(M_1,M_2)=(100,300)$~GeV, $\mu=200~\text{GeV}$.
Also shown is the region excluded by the $B$-physics constraints
and the 0\% contour from the unitarity constraints.}
\end{center}
\end{figure}
%%%%%%%%%%%%%%%%%%%%%%%%%%%%%%%%%%%%%%%%%%%%%%%%%%%%%%%%%%%%%%%%%%%%%%

Next, we show in Fig.~\ref{Fig:delga-mu=200}, the corresponding allowed
regions for $\mu=200~\text{GeV}$. The large-$\tan\beta$ region is then less
constrained by the unitarity constraints, but for the higher value of $M_2$
(right panel) the $R_b$ modification that is caused by neutral-Higgs exchange
starts to exclude high values of $\tan\beta$.

%%%%%%%%%%%%%%%%%%%%%%%%%%%%%%%%%%%%%%%%%%%%%%%%%%%%%%%%%%%%%%%%%%%%%%%%
\subsection{LEP2 non-discovery} 
\label{sect:neutr-LEP-non-discovery}
%%%%%%%%%%%%%%%%%%%%%%%%%%%%%%%%%%%%%%%%%%%%%%%%%%%%%%%%%%%%%%%%%%%%%%%%

The non-discovery of a neutral Higgs boson at LEP2 is relevant only for
$M_1<114.4~\GeV$ \cite{Yao:2006px}.  However, it is well known that these
searches do not exclude certain other light neutral Higgs bosons, if they
couple more weakly to the $Z$ boson, or if they decay to final states that are
more difficult to detect and identify.

The constraint is implemented in an approximate way as follows. Following
\cite{Boonekamp:2004ae} (see also \cite{Achard:2003ty}), we consider the
searches for a neutral Higgs boson that decays to $b\bar b$ jets or a tau
pair. Thus, the quantity of interest is the product of the production cross
section (proportional to the square of the $ZZH_1$ coupling) and the $b \bar
b$ (or $\tau^+\tau^-$) branching ratio (proportional to the square of the $H_1
b\bar b$ coupling).  We denote the reduced sensitivity, as compared to the SM
sensitivity, a ``dilution factor'' $C^2$ \cite{Boonekamp:2004ae}:
\begin{equation}
\sigma_\text{2HDM}(ZH_1\!\to\! Zb\bar b)
=\sigma_\text{SM}(Zh\!\to\! Zb\bar b)
\times C^2(ZH_1\!\to\! Zb\bar b),
\end{equation}
where the dilution factor is the product of a factor $C^2(ZH_1)$ related to
the production and another, $C^2(H_1\to b\bar b)$, related to the branching
ratio:
\begin{equation} \label{Eq:C^2}
C^2(ZH_1\to Zb\bar b)
=C^2(ZH_1)\times C^2(H_1\to b\bar b).
\end{equation}
For the general 2HDM, $C^2(ZH_1\to Zb\bar b)\equiv C^2_\text{2HDM}$ is given
by Eq.~(4.3) in \cite{ElKaffas:2006nt}.  It depends on the neutral-sector
rotation angles, as well as on $\tan\beta$, and is the same for $b\bar b$ and
$\tau^+\tau^-$ final states.

%%%%%%%%%%%%%%%%%%%%%%%%%%%%%%%%%%%%%%%%%%%%%%%%%%%%%%%%%%%%%%%%%%%%%%
\begin{table}[ht]
\begin{center}
\renewcommand{\tabcolsep}{.75em}
\begin{tabular}{|c|c|c|c|c|}
\hline 
$M_1$&80~GeV&100~GeV&114.4~GeV\\
\hline
$b\bar b$ &0.06&0.25&1.0\\
$\tau^+\tau^-$ &0.06&0.2&1.0\\
\hline
\end{tabular} 
\caption{Experimental suppression factors $C^2_\text{exp}$
\cite{Boonekamp:2004ae}.}
\label{tab:C^2}
\end{center}
\end{table}
%%%%%%%%%%%%%%%%%%%%%%%%%%%%%%%%%%%%%%%%%%%%%%%%%%%%%%%%%%%%%%%%%%%%%%

As experimental constraints, we approximate the 95\% C.L.\ bounds obtained in
\cite{Boonekamp:2004ae} by linear interpolations passing through the points
given in Table~\ref{tab:C^2},
and form the {\it ad hoc} single-sided $\chi^2$ penalty (summed over $b\bar b$
and $\tau^+\tau^-$):
\begin{equation}
\chi^2_\text{LEP2} 
=5.99\biggl(\frac{C^2_\text{2HDM}-C^2_\text{exp}}
{1-C^2_\text{exp}}\biggr)^2
\end{equation}
for $ C^2_\text{2HDM}> C^2_\text{exp}$ and $M_1<114.4~\text{GeV}$, and zero
otherwise.  The coefficient 5.99 corresponds to 95\% probability in the
context of our two ``degrees of freedom'', $\tan\beta$ and $M_{H^\pm}$.

%%%%%%%%%%%%%%%%%%%%%%%%%%%%%%%%%%%%%%%%%%%%%%%%%%%%%%%%%%%%%%%%%%%%%%
\begin{figure}[htb]
\vspace*{-2mm}
\begin{center}
\includegraphics[width=95mm]{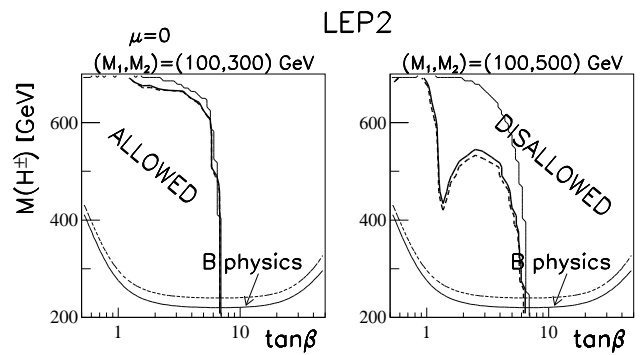}
\vspace*{-10mm}
\caption{\label{Fig:lep2-mu=0}
LEP2 constraint, $\hat\chi^2_\text{LEP2}$.
Left: $(M_1,M_2)=(100,300)$~GeV;
Right: $(M_1,M_2)=(100,500)$~GeV, $\mu=0$
in both cases.
Also shown is the region excluded by the $B$-physics constraints, and the 0\%
contour from Fig.~\ref{Fig:unitarity-mu=0}.}
\end{center}
\end{figure}
%%%%%%%%%%%%%%%%%%%%%%%%%%%%%%%%%%%%%%%%%%%%%%%%%%%%%%%%%%%%%%%%%%%%%%

We show in Fig.~\ref{Fig:lep2-mu=0} the resulting exclusion for
$M_1=100~\text{GeV}$, $\mu=0$ and two values of $M_2$ as indicated.  For the
higher value of $M_2$, we note that some part of the otherwise allowed
parameter space gets excluded. Given that $C^2_\text{2HDM}$ is determined by
$\tan\beta$ and the $\vecalpha$ parameters, on might wonder why this excluded
region depends also on $M_{H^\pm}$. The reason is of course that the subspace
of $\vecalpha$ that is allowed by the positivity and unitarity constraints
depends on $M_{H^\pm}$, and need not overlap with the corresponding subspace
of $\vecalpha$ for which $C^2_\text{2HDM}$ is within the allowed range.
%%%%%%%%%%%%%%%%%%%%%%%%%%%%%%%%%%%%%%%%%%%%%%%%%%%%%%%%%%%%%%%%%%%%%%
\begin{figure}[htb]
\vspace*{-2mm}
\begin{center}
\includegraphics[width=95mm]{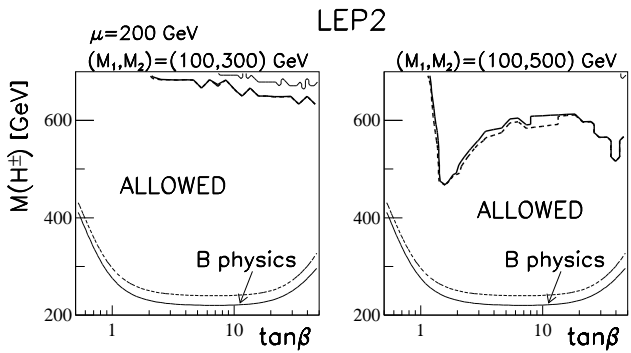}
\vspace*{-10mm}
\caption{\label{Fig:lep2-mu=200}
LEP2 constraint, $\hat\chi^2_\text{LEP2}$.
Left: $(M_1,M_2)=(100,300)$~GeV;
Right: $(M_1,M_2)=(100,500)$~GeV, $\mu=200~\text{GeV}$
in both cases.
Also shown is the region excluded by the $B$-physics constraints, and the 0\%
contour from the unitarity constraints.}
\end{center}
\end{figure}
%%%%%%%%%%%%%%%%%%%%%%%%%%%%%%%%%%%%%%%%%%%%%%%%%%%%%%%%%%%%%%%%%%%%%%

For $\mu=200~\text{GeV}$, the unitarity constraints no longer exclude high
values of $\tan\beta$, but we see that the LEP non-discovery excludes high
values of charged-Higgs mass (via its impact on the rotation matrix).
In fact, at large values of $\tan\beta$, if $M_1<114.4~\text{GeV}$,
the LEP2 non-discovery requires (see Eq.~(4.3) of \cite{ElKaffas:2006nt})
\begin{equation} \label{Eq:LEP2-constraint}
\frac{1}{\cos^2\beta}\, R_{12}^2\, [R_{11}^2+R_{13}^2]<1.
\end{equation}
For small values of $\cos\beta$, this is satisfied in three separate regions
[see Eq.~(\ref{Eq:R-matrix}) and Fig.~\ref{Fig:alphas-conthi}]:
\begin{align} \label{Eq:LEP2-cases}
&(i):\quad &\alpha_1&\simeq0, \nonumber \\
&(ii):\quad &|\alpha_2|&\simeq\pi/2, \nonumber \\
&(iii):\quad &|\alpha_1|&\simeq\pi/2\quad\text{and}\quad
\alpha_2\simeq0.
\end{align}
%%%%%%%%%%%%%%%%%%%%%%%%%%%%%%%%%%%%%%%%%%%%%%%%%%%%%%%%%%%%%%%%%%%%%%%%
\subsection{The $\rho$ parameter} 
\label{sect:neutr-Deltarho}
%%%%%%%%%%%%%%%%%%%%%%%%%%%%%%%%%%%%%%%%%%%%%%%%%%%%%%%%%%%%%%%%%%%%%%%%

The $\rho$ parameter, defined as
\begin{equation}
\rho=\frac{M_W^2}{M_Z^2\cos^2\theta_\text{W}},
\end{equation}
is very sensitive to fields that couple to the $W$ and $Z$
\cite{Ross:1975fq}. 
Experimentally, $\rho$ is constrained as \cite{:2005em}
\begin{equation}
\rho_\text{exp}=1.0050\pm0.0010.
\end{equation}
The deviation from unity is mostly due to the top-quark one-loop contributions,
but there is also a weak dependence on the SM Higgs mass.
In order to extract this quantity from the data, one fits for a SM Higgs mass,
the contribution of which should then be subtracted from the 2HDM prediction.

The additional Higgs fields of the 2HDM can easily spoil
the agreement with the SM \cite{Bertolini:1985ia}.  Roughly speaking, this
constraint requires the Higgs masses to be not too far from the $W$ and $Z$
masses, and not very much apart.  For the general CP-non-conserving 2HDM, the
results of \cite{Bertolini:1985ia} for the contributions to $\rho$, denoted
$\Delta\rho=\rho_\text{2HDM}-\rho_\text{SM}$, were generalized in
\cite{ElKaffas:2006nt} and given by Eqs.~(4.8)--(4.12) there.

%%%%%%%%%%%%%%%%%%%%%%%%%%%%%%%%%%%%%%%%%%%%%%%%%%%%%%%%%%%%%%%%%%%%%%
\begin{figure}[htb]
\vspace*{-2mm}
\begin{center}
\includegraphics[width=95mm]{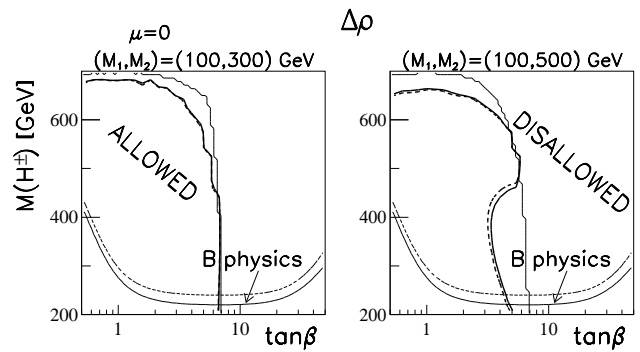}
\vspace*{-10mm}
\caption{\label{Fig:rho-mu=0} Exclusions due to the $\Delta\rho$
constraint, $\hat\chi_\rho^2$, for $(M_1,M_2)=(100,300)$~GeV, 
and $(M_1,M_2)=(100,500)$~GeV, $\mu=0$ in both cases.
Also shown is the region excluded by the $B$-physics constraints, and the 0\%
contour from Fig.~\ref{Fig:unitarity-mu=0}.}
\end{center}
\end{figure}
%%%%%%%%%%%%%%%%%%%%%%%%%%%%%%%%%%%%%%%%%%%%%%%%%%%%%%%%%%%%%%%%%%%%%%

In order to study this constraint, we evaluate $\chi_\rho^2$ as defined by
\begin{equation}
\chi^2_\rho
=\biggl(\frac{\Delta\rho_\text{2HDM}}{\sigma(\rho)}\biggr)^2, 
\label{Eq:chi2-rho}\\
\end{equation}
with $\sigma(\rho)=0.0010$, subtracting the SM contribution
corresponding to an SM Higgs mass $M_0=129~\text{GeV}$.

\begin{figure}[htb]
\vspace*{-2mm}
\begin{center}
\includegraphics[width=95mm]{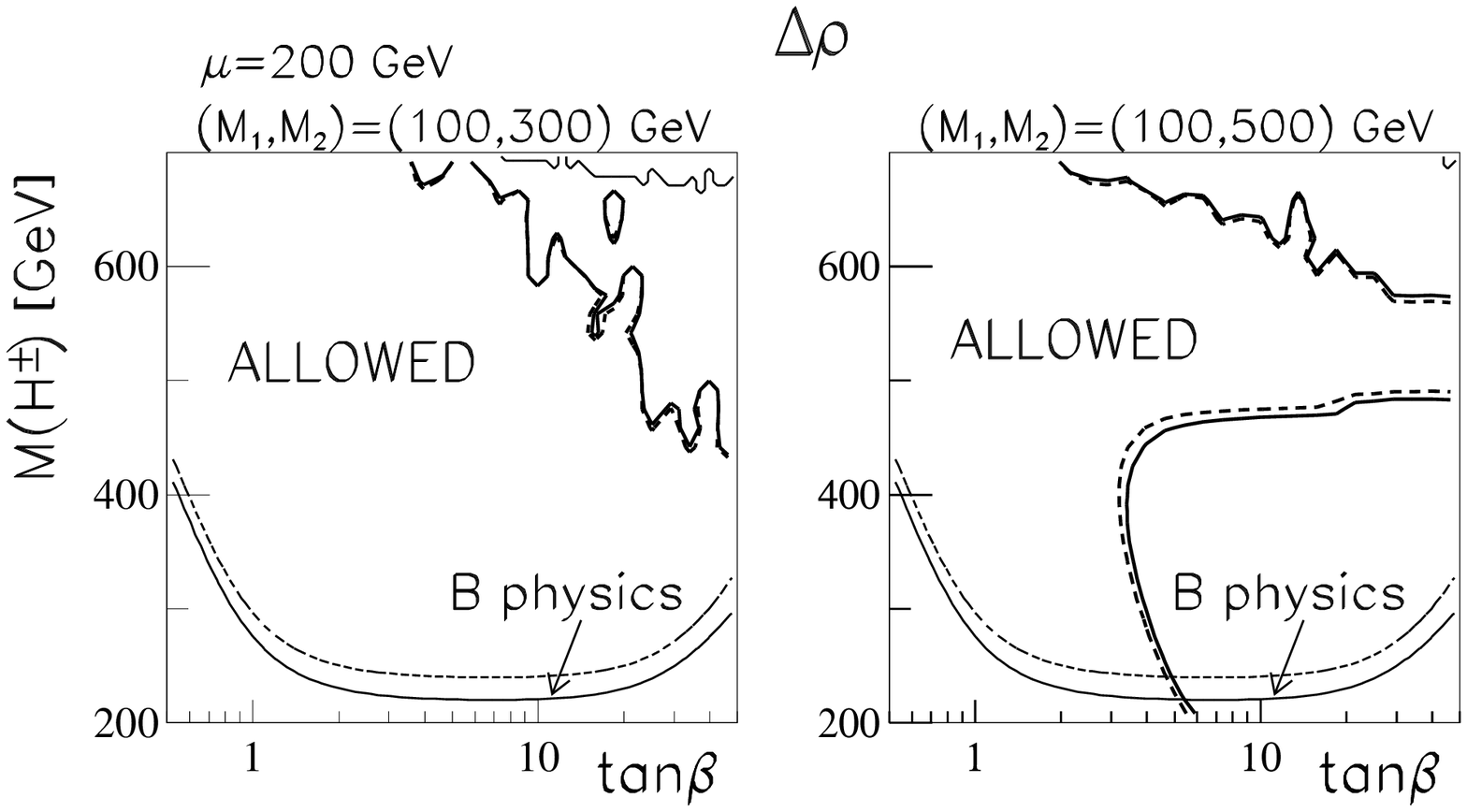}
\vspace*{-10mm}
\caption{\label{Fig:rho-mu=200} Exclusions due to the $\Delta\rho$
constraint, $\hat\chi_\rho^2$, for $(M_1,M_2)=(100,300)$~GeV, 
and $(M_1,M_2)=(100,500)$~GeV, $\mu=200~\text{GeV}$ in both cases.
Also shown is the region excluded by the $B$-physics constraints, and the 0\%
contour from the unitarity constraint.}
\end{center}
\end{figure}

In Figs.~\ref{Fig:rho-mu=0} and \ref{Fig:rho-mu=200} we show the impact of
$\hat\chi_\rho^2$ in constraining the parameter space for $\mu=0$ and
$\mu=200~\text{GeV}$, respectively.  At the higher value of $M_2$, and high
values of $\tan\beta$, this constraint tends to exclude both low and high
values of $M_{H^\pm}$. The irregular boundaries seen in
Fig.~\ref{Fig:rho-mu=200} are obviously due to the scanning not finding the
whole allowed region. (At high $\mu$ and high $\tan\beta$, only very tiny
regions in $\alpha_1$ and $\alpha_2$ are allowed \cite{Kaffas:2007rq}.)
%%%%%%%%%%%%%%%%%%%%%%%%%%%%%%%%%%%%%%%%%%%%%%%%%%%%%%%%%%%%%%%%%%%%%%
%%%%%%%%%%%%%%%%%%%%%%%%%%%%%%%%%%%%%%%%%%%%%%%%%%%%%%%%%%%%%%%%%%%%%%

It is instructive to consider the contribution to $\Delta\rho$ in the limit of
large values of $\tan\beta$. The expressions (4.10) and (4.12) of
\cite{ElKaffas:2006nt} simplify considerably in the limit $\sin\beta\to1$,
and provide an understanding of features seen in Figs.~\ref{Fig:rho-mu=0}
and \ref{Fig:rho-mu=200}:
\begin{align} \label{Eq:delta-rho-mh-ch}
&A_{WW}^{HH}(0)-\cos^2\theta_W\, A_{ZZ}^{HH}(0)\nonumber\\
&\to\frac{g^2}{64\pi^2}\sum_j\Bigl[
(R^2_{j1} +R^2_{j3})
F_{\Delta\rho}(M_{H^\pm}^2,M_j^2)
\nonumber \\
&\quad -\sum_{k>j}(R_{j1}R_{k3}-R_{k1}R_{j3})^2
F_{\Delta\rho}(M_j^2,M_k^2) \Bigr],
\end{align}
and
\begin{align} \label{Eq:delta-rho-rem}
&A_{WW}^{HG}(0)-\cos^2\theta_W\, A_{ZZ}^{HG}(0)\nonumber\\
&\to\frac{g^2}{64\pi^2}\Bigl[\sum_j R^2_{j2} \,
\Bigl(3F_{\Delta\rho}(M_Z^2,M_j^2)-3F_{\Delta\rho}(M_W^2,M_j^2)\Bigl) 
\nonumber \\
&\quad
+3F_{\Delta\rho}(M_W^2,M_0^2)
-3F_{\Delta\rho}(M_Z^2,M_0^2)\Bigr]
\end{align}
with $F_{\Delta\rho}(m_1^2,m_2^2)$ found in Eq.~(4.11) of
\cite{ElKaffas:2006nt}. This function vanishes when the two masses are equal,
and grows quadratically with the bigger of the two masses.  
The contribution given by (\ref{Eq:delta-rho-rem}) is rather small,
since $M_Z$ and $M_W$ are relatively close.

The only part sensitive to the charged-Higgs mass is the first sum in
(\ref{Eq:delta-rho-mh-ch}), which may however get significant contributions
from all three neutral Higgs bosons, $j=1,2,3$. Consider first the
contribution from $H_1$.  Barring cancellations (see below), a viable
parameter point must for large $\tan\beta$ and large $M^2_{H^\pm}\gg M_1$ have
\begin{equation} \label{Eq:small-factor}
R^2_{11} +R^2_{13}\ll1.
\end{equation}
Expressed in terms of the angles, this means that $c_1^2c_2^2+s_2^2\ll1$, or
\begin{equation} \label{Eq:no-delta-rho}
|\alpha_1|\sim\pi/2, \quad \text{and} \ 
\alpha_2\sim0.
\end{equation}
Note that this condition is compatible with (\ref{Eq:LEP2-constraint}) and
corresponds to case $(iii)$ of (\ref{Eq:LEP2-cases}).

Next, we consider the case
(with increasing splitting between $M_2$ and $M_3$, the
contribution to $\Delta\rho$ will increase)
\begin{equation}
M_{H^\pm}\ll M_2\simeq M_3,
\end{equation}
then again barring cancellations, the $\Delta\rho$ condition requires
\begin{equation}
\sum_{j=2}^3(R_{j1}^2+R_{j3}^2)=c_1^2s_2^2+s_1^2+c_2^2\ll1.
\end{equation}
This condition {\it cannot} be satisfied, and low values of $M_{H^\pm}$ are
thus excluded when $\tan\beta$ is large and $M_2$ is large.

For high values of $M_{H^\pm}$, the most sensitive contribution is that
involving $H_1$.  But unless $M_{H^\pm}$ is close to $M_2$, the condition
(\ref{Eq:no-delta-rho}), related to the $H_1$ contribution, is not sufficient.
The $H_2$ and $H_3$ contributions are proportional to $R_{21}^2+R_{23}^2$ and
$R_{31}^2+R_{33}^2$, respectively.  In the limit (\ref{Eq:no-delta-rho}),
these are both equal to 1, i.e., when the rotation matrix is adjusted such as
to cancel the $H_1$ contribution to the first sum in
(\ref{Eq:delta-rho-mh-ch}), there is no suppression of the $H_2$ and $H_3$
contributions.  As a result, such high values of $M_{H^\pm}$ are forbidden.

An exception to this situation arises when $\mu$ is large compared with $M_2$.
Then, there can be a considerable splitting between $M_2$ and $M_3$,
$M_{H^\pm}$ and $M_3$ can be similar, and a cancellation between
the ($M_1$, $M_{H^\pm}$) and the  ($M_1$, $M_3$) terms 
of (\ref{Eq:delta-rho-mh-ch}) is possible.
As a result, for large values of $\mu$, large values of $M_{H^\pm}$
can be allowed.

Finally, we consider the situation
\begin{equation}
M_1\ll M_2\simeq M_3\simeq M_{H^\pm},
\end{equation}
which applies to the ``finger'' protruding to the right in the right panel of
Fig.~\ref{Fig:rho-mu=200}. Then, the contributions of the two sums in
Eq.~(\ref{Eq:delta-rho-mh-ch}) tend to cancel. For $j=1$ and $k=2,3$, we get
\begin{align}
\bigl[&R_{11}^2+R_{13}^2-(R_{11}R_{23}-R_{21}R_{13})^2 \nonumber \\
&-(R_{11}R_{33}-R_{31}R_{13})^2\bigr]
 F_{\Delta\rho}(M_1^2,M_2^2)=0,
\end{align}
where we have used the orthogonality of the rotation matrix.  Thus, when
$\tan\beta$ is large, and $M_2\simeq M_3\simeq M_{H^\pm}$, there is a
cancellation among the terms in (\ref{Eq:delta-rho-mh-ch}), only the (small)
(\ref{Eq:delta-rho-rem}) part of the $\Delta\rho$ constraint is relevant.
%%%%%%%%%%%%%%%%%%%%%%%%%%%%%%%%%%%%%%%%%%%%%%%%%%%%%%%%%%%%%%%%%%%%%%%%
\subsection{Muon anomalous magnetic moment $a_\mu$} 
\label{sect:neutr-g-2}
%%%%%%%%%%%%%%%%%%%%%%%%%%%%%%%%%%%%%%%%%%%%%%%%%%%%%%%%%%%%%%%%%%%%%%%%

The precisely measured muon anomalous magnetic moment \cite{Bennett:2004pv},
\begin{equation}
a_{\mu,\text{exp}}\equiv\half(g\!-\!2)_{\mu,\text{exp}}
=11659208(5.4)(3.3)\times10^{-10}
\end{equation}
is a sensitive probe of new physics (for a recent review, see
\cite{Jegerlehner:2007xe}).  The statistical and systematic uncertainties
(given in parentheses) combine to an over-all uncertainty of
$6.3\times10^{-10}$.  The corresponding SM prediction, including weak and
strong effects, is \cite{Jegerlehner:2007xe}
\begin{equation}
a_{\mu,\text{SM}}
=11659179.3(6.8)\times10^{-10},
\end{equation}
creating a $3\sigma$ ``tension'' with the experimental result.

In the 2HDM, there are additional contributions, dominated by the two-loop
Barr--Zee effect \cite{Barr:1990vd} with a photon and a Higgs field connected
to a heavy-fermion loop.  For the CP-conserving case, the contribution is
given by \cite{Chang:2000ii,Cheung:2003pw}.  For the general (CP-violating)
2HDM, the top-quark contributions to $a_\mu$ for the muon, is given by
Eq.~(4.13) in \cite{ElKaffas:2006nt}, whereas the $b$-quark contributions to
the fermion loop is given by
\begin{align} \label{Eq:Delta-amu}
\Delta a_\mu=\frac{N_c \alpha_\text{e.m.}}{4\pi^3 v^2}\,
m_\mu^2 Q_b^2
\sum_j\biggl[&\tan^2\beta R_{j3}^2\, g\biggl(\frac{m_b^2}{M_j^2}\biggr) 
\nonumber \\
-&\frac{1}{\cos^2\beta}R_{j1}^2\, 
f\biggl(\frac{m_b^2}{M_j^2}\biggr)\biggr],
\end{align}
with $N_c=3$ the number of colours associated with the fermion loop,
$\alpha_\text{e.m.}$ the electromagnetic finestructure constant, $Q_b=-1/3$ and
$m_b$ the $b$-quark charge and mass, and $m_\mu$ the muon mass.  
The $\tau$-loop contribution, which we also include, is given by a similar
expression, with obvious substitutions for the colour factor, charge and mass.
The functions $f$ and $g$ are given in \cite{Barr:1990vd}. 

At low values of $\tan\beta$, these contributions are negligible, but the $b$-
and $\tau$-loop contributions can become relevant at very large values of
$\tan\beta$.  As a measure of the possible conflict with the 2HDM, we consider
\begin{equation}
\chi^2_{a_\mu}
=\left(\frac{\Delta a_{\mu,\text{2HDM}}}
{\sigma(a_\mu)}\right)^2,
\end{equation}
where $\Delta a_{\mu,\text{2HDM}}$ is the 2HDM-specific contribution, and for
the uncertainty we take the (SM) theoretical value,
$\sigma(a_\mu)=6.8\times10^{-10}$, since this is larger than the experimental
one.

Actually, this constraint does not have any significant impact within the
range of $\tan\beta$ considered.  Let us consider its ``natural value'' as
that contributed by one of the two terms in (\ref{Eq:Delta-amu}), with the
rotation matrix element set to 1. This reaches $\chi^2={\cal O}(1)$ for
$\tan\beta$ of the order of 70. However, there can be significant reductions
by the rotation matrix elements, and also cancellations among the two terms.
At such high values of $\tan\beta$ the $B\to\tau\nu_\tau$ constraint (see
Sec.~\ref{Sec:B-tau-nu}) is important at low values of $M_{H^\pm}$, and the
unitarity constraint may be important at high $M_{H^\pm}$ (depending on the
relative magnitude of $M_2$ and $\mu$).

We recall that at high $\tan\beta$, the rotation matrix is rather constrained
by unitarity \cite{Kaffas:2007rq}. Let us focus on the contribution of the
lightest neutral Higgs boson, $H_1$, whose contributions to
(\ref{Eq:Delta-amu}) are given by $R_{11}^2$ and $R_{13}^2$. Thus, in spite of
the enhancement given by the $\tan\beta$-dependent factors, this contribution
to $a_\mu$ may be small if
\begin{equation}
|\alpha_1|\sim\pi/2, \quad \text{and} \ 
\alpha_2\sim0.
\end{equation}
In the limit of large $\tan\beta$ and large $M_{H^\pm}$, this is actually the
only region allowed by the $\Delta\rho$ constraint, see
Eq.~(\ref{Eq:no-delta-rho}). We conclude that at large $\tan\beta$ and large
$M_{H^\pm}$, the $a_\mu$ constraint is covered by the $\Delta\rho$ constraint.
But this coefficient $R_{11}^2+R_{13}^2$ arises in one case from the Yukawa
couplings, and in the other from the gauge--Higgs couplings.  Furthermore, at
large $\tan\beta$ and moderate $M_{H^\pm}$ values, the $a_\mu$ constraint is
covered by the $B\to\tau\nu_\tau$ constraint.
%%%%%%%%%%%%%%%%%%%%%%%%%%%%%%%%%%%%%%%%%%%%%%%%%%%%%%%%%%%%%%%%%%%%%%%%
\subsection{Summary on neutral-sector constraints} 
\label{sect:neutr-summary}
%%%%%%%%%%%%%%%%%%%%%%%%%%%%%%%%%%%%%%%%%%%%%%%%%%%%%%%%%%%%%%%%%%%%%%%%

We will here summarize the conclusions on the neutral-sector constraints,
treating first the simpler case of $\mu=0$ (where $\tan\beta$ is bounded), and
next comment on the less restrictive case of ``large'' $\mu$ (where also
larger values of $\tan\beta$ are allowed).  The $R_b$ constraint is at low
values of $\tan\beta$ dominated by the charged-Higgs-exchange
contribution. This part of the $R_b$ constraint is thus independent of the
neutral sector.

%%%%%%%%%%%%%%%%%%%%%%%%%%%%%%%%%%%%%%%%%%%%%%%%%%%%%%%%%%%%%%%%%%%%%%%%
\subsubsection{The case $\mu=0$} 
%%%%%%%%%%%%%%%%%%%%%%%%%%%%%%%%%%%%%%%%%%%%%%%%%%%%%%%%%%%%%%%%%%%%%%%%

For $\mu=0$, and $(M_1,M_2)=(100,300)~\text{GeV}$, the 
only neutral-sector constraint that has some impact, apart from
$R_b$ at low values of $\tan\beta$, is
the $\Delta\rho$ constraint, which excludes the higher range of $M_{H^\pm}$,
as illustrated in the left panel of Fig.~\ref{Fig:rho-mu=0}.
However, for $M_2=500~\text{GeV}$, other constraints are also important.
The LEP2 non-discovery 
rules out large values of $\tan\beta$ and $M_{H^\pm}$, and
to some extent, also the $\Delta\rho$ constraint rules out some
region of large $\tan\beta$.

However, with $\mu=0$, high values of $\tan\beta$ are also excluded by the
unitarity constraints, and, to some extent, the low values of $M_{H^\pm}$ 
are excluded by the $B$-physics constraints.

%%%%%%%%%%%%%%%%%%%%%%%%%%%%%%%%%%%%%%%%%%%%%%%%%%%%%%%%%%%%%%%%%%%%%%%%
\subsubsection{The case $\mu>M_1$} 
%%%%%%%%%%%%%%%%%%%%%%%%%%%%%%%%%%%%%%%%%%%%%%%%%%%%%%%%%%%%%%%%%%%%%%%%

When $\mu>M_1$, large values of $\tan\beta$ become accessible. 
This parameter region is known as the decoupling region \cite{Gunion:2002zf}.
We here distinguish two cases
\begin{alignat}{2}
(i)& &\quad &M_1<\mu<M_2,  \nonumber \\
(ii)& &\quad &M_1<M_2<\mu. 
\end{alignat}

The $R_b$ constraint, which at low $\tan\beta$ is dominated
by the charged-Higgs contributions, can at large $\tan\beta$ also exclude
some region of neutral Higgs boson mass values (compare the left and right 
panels of Fig.~\ref{Fig:delga-mu=200}).
Furthermore, the LEP2 constraint may exclude high values
of $M_{H^\pm}$ (see right panel of Fig.~\ref{Fig:lep2-mu=200})
and the $\Delta\rho$ constraint may at high $\tan\beta$
constrain the range of $M_{H^\pm}$ values to a band
around $M_2$.

%%%%%%%%%%%%%%%%%%%%%%%%%%%%%%%%%%%%%%%%%%%%%%%%%%%%%%%%%%%%%%%%%%%%%%%%
\section{Combining all constraints} 
\setcounter{equation}{0}
\label{sect:combining}
%%%%%%%%%%%%%%%%%%%%%%%%%%%%%%%%%%%%%%%%%%%%%%%%%%%%%%%%%%%%%%%%%%%%%%%%

%%%%%%%%%%%%%%%%%%%%%%%%%%%%%%%%%%%%%%%%%%%%%%%%%%%%%%%%%%%%%%%%%%%%%%
Let us now combine all constraints.  This is done by a dedicated scan over
$\hat\vecalpha$ for each point in $\tan\beta$ and $M_{H^\pm}$. The value of
$\chi^2$ is determined as
\begin{equation} \label{Eq:chi2-all}
\chi^2=\chi^2_\text{general}+\sum_i\chi_i^2,
\end{equation}
where $\chi^2_\text{general}$ is given by Eq.~(\ref{Eq:chi2-general}) and the
sum runs over the observables $R_b$, LEP2 non-discovery, $\Delta\rho$ and
$a_\mu$, all of them evaluated at the same point in $\hat\vecalpha$.  This
quantity is then minimized over $\hat\vecalpha$:
\begin{equation}
\hat\chi^2=\mathop{\min}_{\hat\vecalpha\in\vecalpha_+}\chi^2,
\end{equation}
for fixed $\tan\beta$ and $M_{H^\pm}$ and allowed regions are determined.  In
the $\tan\beta$--$M_{H^\pm}$ plane, the allowed regions will in general be
less that the intersection of the regions that are allowed by the individual
constraints.  The reason is that the individual constraints may refer to
different parts of the three-dimensional $\vecalpha$ space.

We shall split this discussion into the two cases $\mu=0$ and $M_1<\mu$.  In
the former case, unitarity restricts the allowed range of $\tan\beta$, as
illustrated in Fig.~\ref{Fig:unitarity-mu=0}, whereas in the latter case also
higher values of $\tan\beta$ are allowed.
%%%%%%%%%%%%%%%%%%%%%%%%%%%%%%%%%%%%%%%%%%%%%%%%%%%%%%%%%%%%%%%%%%%%%%%%
\subsection{Combining all constraints for $\mu=0$} 
%%%%%%%%%%%%%%%%%%%%%%%%%%%%%%%%%%%%%%%%%%%%%%%%%%%%%%%%%%%%%%%%%%%%%%%%

\begin{figure}[htb]
\vspace*{-2mm}
\begin{center}
\includegraphics[width=95mm]{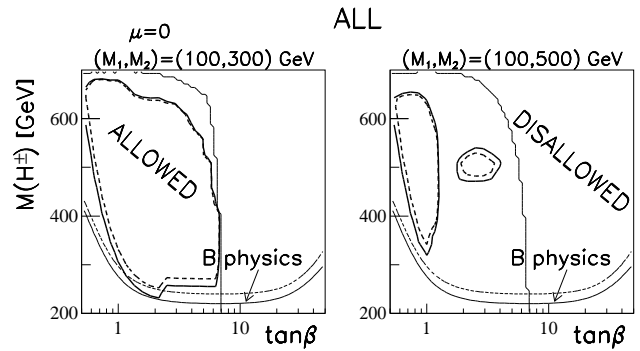}
\vspace*{-10mm}
\caption{\label{Fig:neu-all-mu=0} Exclusions due to all constraints, for
$(M_1,M_2)=(100,300)$~GeV, and $(M_1,M_2)=(100,500)$~GeV, both with $\mu=0$.
Heavy dashed: 90\% C.L., heavy solid: 95\% C.L.  Also shown is the region
excluded by the $B$-physics constraints (thin dashed and solid), and the 0\%
contour from Fig.~\ref{Fig:unitarity-mu=0} (thin solid).}
\end{center}
\end{figure}

In Fig.~\ref{Fig:neu-all-mu=0} we display the 90 and 95\% C.L.\ limits for
$(M_1,M_2)=(100,300)$~GeV and $(100,500)$~GeV, in both cases for $\mu=0$.  For
the case of moderately low $M_2=300~\text{GeV}$, we note that there is little
additional exclusion, other than that due to the $B$-physics constraints and
unitarity. The little extra is due to the $\Delta\rho$ constraint, at high
values of $M_{H^\pm}$.  However, for $M_2=500~\text{GeV}$, there is a
considerable reduction of the allowed parameter space at
$\tan\beta\gsim1-1.5$.  In this range of $\tan\beta$ values, we see from
Fig.~\ref{Fig:lep2-mu=0} (right panel), that high values of $M_{H^\pm}$ are
excluded by the LEP2 non-discovery constraint, and from
Fig.~\ref{Fig:rho-mu=0} (right panel), we see that low values of $M_{H^\pm}$
are excluded by the $\Delta\rho$ constraint.

\begin{figure}[htb]
\vspace*{-2mm}
\begin{center}
\includegraphics[width=95mm]{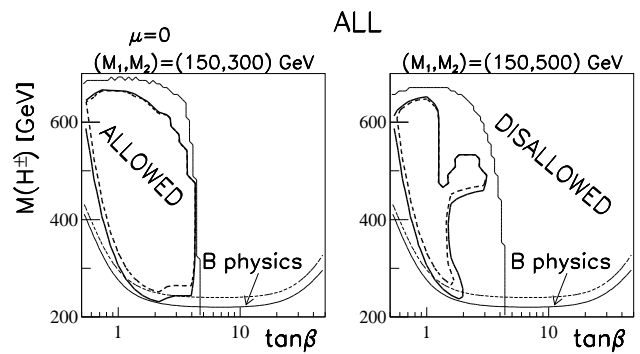}
\vspace*{-10mm}
\caption{\label{Fig:neu-all-150-mu=0} Similar to Fig.~\ref{Fig:neu-all-mu=0},
for $M_1=150~\text{GeV}$.}
\end{center}
\end{figure}

Similarly, Fig.~\ref{Fig:neu-all-150-mu=0} is devoted to the case
$M_1=150~\text{GeV}$. In this case, the LEP2 non-discovery plays no
role, but there are of course also other differences, due to the way different
parameters are correlated by the constraints.
Overall, this case is less constrained than the $M_1=100~\text{GeV}$ case of
Fig.~\ref{Fig:rho-mu=200}.
%%%%%%%%%%%%%%%%%%%%%%%%%%%%%%%%%%%%%%%%%%%%%%%%%%%%%%%%%%%%%%%%%%%%%%
%%%%%%%%%%%%%%%%%%%%%%%%%%%%%%%%%%%%%%%%%%%%%%%%%%%%%%%%%%%%%%%%%%%%%%

%%%%%%%%%%%%%%%%%%%%%%%%%%%%%%%%%%%%%%%%%%%%%%%%%%%%%%%%%%%%%%%%%%%%%%
%%%%%%%%%%%%%%%%%%%%%%%%%%%%%%%%%%%%%%%%%%%%%%%%%%%%%%%%%%%%%%%%%%%%%%

%%%%%%%%%%%%%%%%%%%%%%%%%%%%%%%%%%%%%%%%%%%%%%%%%%%%%%%%%%%%%%%%%%%%%%%%
\subsection{Combining all constraints for $M_1<\mu$} 
%%%%%%%%%%%%%%%%%%%%%%%%%%%%%%%%%%%%%%%%%%%%%%%%%%%%%%%%%%%%%%%%%%%%%%%%

\begin{figure}[htb]
\vspace*{-2mm}
\begin{center}
\includegraphics[width=95mm]{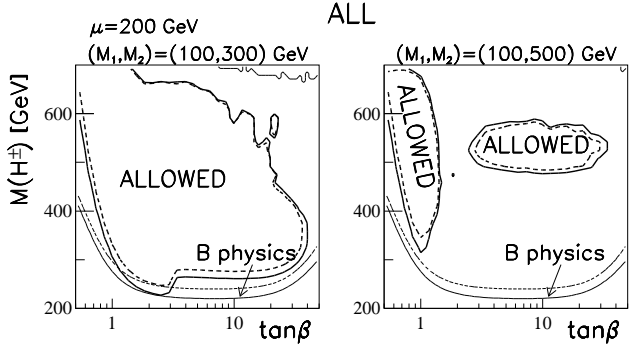}
\vspace*{-10mm}
\caption{\label{Fig:neu-all-mu=200} Similar to Fig.~\ref{Fig:neu-all-mu=0},
for $\mu=200$.}
\end{center}
\end{figure}

\begin{figure}[htb]
\vspace*{-2mm}
\begin{center}
\includegraphics[width=95mm]{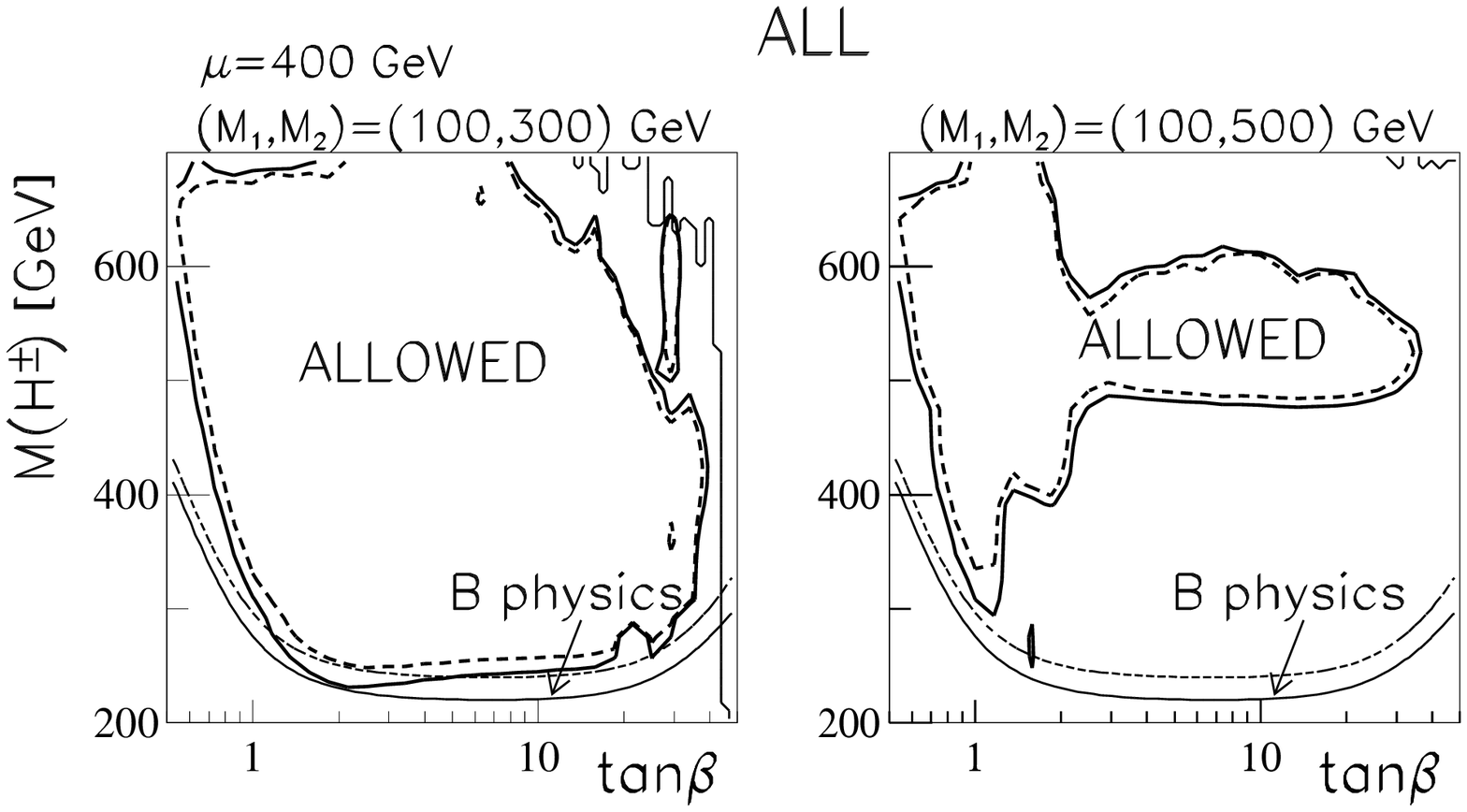}
\vspace*{-10mm}
\caption{\label{Fig:neu-all-mu=400} Similar to Fig.~\ref{Fig:neu-all-mu=0},
for $\mu=400$.}
\end{center}
\end{figure}

\begin{figure}[htb]
\vspace*{-2mm}
\begin{center}
\includegraphics[width=95mm]{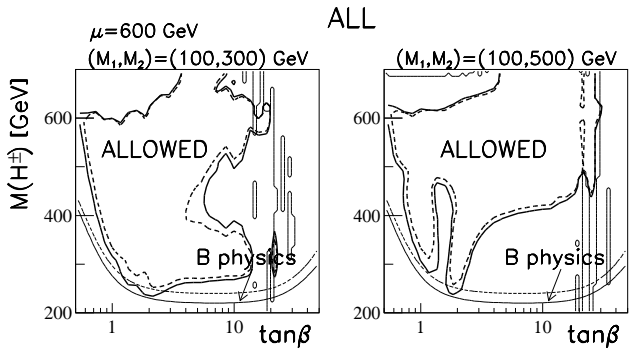}
\vspace*{-10mm}
\caption{\label{Fig:neu-all-mu=600} Similar to Fig.~\ref{Fig:neu-all-mu=0},
for $\mu=600$.}
\end{center}
\end{figure}

When $M_1<\mu$, the unitarity constraints no longer restrict $\tan\beta$ to
low and moderate values. However, we shall see that various other constraints
may cause a cut-off for large $\tan\beta$. For $M_1=100~\text{GeV}$ and two
values of $M_2$, namely 300~GeV and 500~GeV, we display in
Figs.~\ref{Fig:neu-all-mu=200}--\ref{Fig:neu-all-mu=600} the allowed regions
for a range of $\mu$-values, from 200~GeV to 600~GeV.

%%%%%%%%%%%%%%%%%%%%%%%%%%%%%%%%%%%%%%%%%%%%%%%%%%%%%%%%%%%%%%%%%%%%%%
%%%%%%%%%%%%%%%%%%%%%%%%%%%%%%%%%%%%%%%%%%%%%%%%%%%%%%%%%%%%%%%%%%%%%%
%%%%%%%%%%%%%%%%%%%%%%%%%%%%%%%%%%%%%%%%%%%%%%%%%%%%%%%%%%%%%%%%%%%%%%
%%%%%%%%%%%%%%%%%%%%%%%%%%%%%%%%%%%%%%%%%%%%%%%%%%%%%%%%%%%%%%%%%%%%%%
%%%%%%%%%%%%%%%%%%%%%%%%%%%%%%%%%%%%%%%%%%%%%%%%%%%%%%%%%%%%%%%%%%%%%%
%%%%%%%%%%%%%%%%%%%%%%%%%%%%%%%%%%%%%%%%%%%%%%%%%%%%%%%%%%%%%%%%%%%%%%

For the lower value, $M_2=300~\text{GeV}$ (left panels), the allowed region in
the $\tan\beta$--$M_{H^\pm}$ plane is fairly extended, whereas for the higher
value, $M_2=500~\text{GeV}$, it is more constrained, until $\mu$ reaches
values comparable to $M_2$ (see the right panels of
Figs.~\ref{Fig:neu-all-mu=200}, \ref{Fig:neu-all-mu=400} and
\ref{Fig:neu-all-mu=600}).

For $\mu=500~\text{GeV}$ and $M_2=500~\text{GeV}$ (not shown), the exclusion
of low and high values of $M_{H^\pm}$ for $\tan\beta\gsim2$ is due to the
$\Delta\rho$ constraint, whereas the LEP2 non-discovery by itself does not
exclude anything in this case. However, the simultaneous imposition of the
LEP2 non-discovery and the $\Delta\rho$ constraints yields a ``forbidden
finger'' at $\tan\beta\sim1.5$ and low $M_{H^\pm}$, as well as some exclusion
at low $\tan\beta$ and high $M_{H^\pm}$.

In all cases, we note that simultaneously high values of both $\tan\beta$ and
$M_{H^\pm}$ are excluded, except when $M_2<\mu$. This is due to the
$\Delta\rho$ constraint, as discussed in Sec.~\ref{sect:neutr-Deltarho}. For
the cases shown ($M_1=100~\text{GeV}$), also the LEP2 non-discovery plays a
role, but this will of course not have any impact for $M_1>114.4~\text{GeV}$.

%%%%%%%%%%%%%%%%%%%%%%%%%%%%%%%%%%%%%%%%%%%%%%%%%%%%%%%%%%%%%%%%%%%%%%%%
\section{Profile of surviving parameter space}
\label{sect:profile}
\setcounter{equation}{0}
%%%%%%%%%%%%%%%%%%%%%%%%%%%%%%%%%%%%%%%%%%%%%%%%%%%%%%%%%%%%%%%%%%%%%%%%

We shall here give a profile of the surviving parameter space in terms
of three ``hidden'' parameters, $\alpha_1$, $\alpha_2$ and $M_3$.
In discussing the surviving parameter space, we shall distinguish between low
and high values of $\tan\beta$, giving some details relevant to $\tan\beta<5$
and $\tan\beta>10$.

%%%%%%%%%%%%%%%%%%%%%%%%%%%%%%%%%%%%%%%%%%%%%%%%%%%%%%%%%%%%%%%%%%%%%%%%
\subsection{Low values of $\tan\beta$} 
%%%%%%%%%%%%%%%%%%%%%%%%%%%%%%%%%%%%%%%%%%%%%%%%%%%%%%%%%%%%%%%%%%%%%%%%

\begin{figure}[htb]
\vspace*{-2mm}
\begin{center}
\includegraphics[width=95mm]{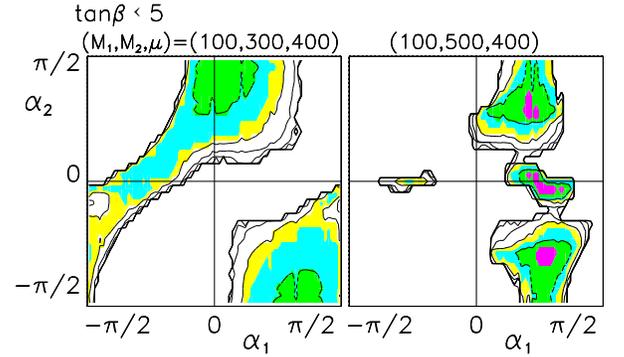}
\vspace*{-10mm}
\caption{\label{Fig:alphas-contlo} Normalized distributions of allowed regions
in the $\alpha_1$--$\alpha_2$ space, for low values of $\tan\beta$ values.
Contours are shown at each negative power of 10, as appropriate.  Yellow
(light blue) indicates where the normalized distribution is higher than
$10^{-4}$ ($3\times10^{-4}$); green (purple) levels above $10^{-3}$
($3\times10^{-3}$). Along the lines $\alpha_2=\pm\pi/2$, $H_1$ is CP-odd, and
there is no CP violation.}
\end{center}
\end{figure}

At low values of $\tan\beta$, both low and high values of $\mu$ lead to
consistent solutions, with values of $\alpha_1$ and $\alpha_2$ distributed
over extended regions of the parameter space, as shown in
Fig.~\ref{Fig:alphas-contlo} for the case $\mu=400~\text{GeV}$,
$M_1=100~\text{GeV}$ and two values of $M_2$, namely 300~GeV and 500~GeV.
(Further plots of this kind, but taking into account {\it only} the positivity
and unitarity constraints, are presented in \cite{Kaffas:2007rq}.)

We have here plotted the distributions of {\it all} $\alpha_1$ and $\alpha_2$
for which the total $\chi^2<5.99$ (see Eq.~(\ref{Eq:chi2-all})), i.e., in
general several points in $(\alpha_1,\alpha_2)$ for each point in
$(\tan\beta,M_{H^\pm})$.

%%%%%%%%%%%%%%%%%%%%%%%%%%%%%%%%%%%%%%%%%%%%%%%%%%%%%%%%%%%%%%%%%%%%%%
%%%%%%%%%%%%%%%%%%%%%%%%%%%%%%%%%%%%%%%%%%%%%%%%%%%%%%%%%%%%%%%%%%%%%%

At very low $\tan\beta$, an important constraint is the positivity of
$\lambda_2$:
\begin{align}
\lambda_2&=\frac{1}{s_\beta^2v^2} \label{Eq:lambda2}
[s_1^2c_2^2M_1^2
+(c_1c_3-s_1s_2s_3)^2M_2^2 \nonumber \\
&+(c_1s_3+s_1s_2c_3)^2M_3^2-c_\beta^2\mu^2]>0, 
\end{align}
and the constraint from unitarity that it does not become ``large''.  For
small $\mu$, the $M_3^2$-term cannot be too large (in order not to violate
unitarity). This requires
\begin{equation}
|c_1s_3+s_1s_2c_3|\ll1.
\end{equation}
As $\mu$ becomes large, with $M_2$ fixed, the $M_3^2$-term must compensate the
$\mu^2$-term, with $|c_1s_3+s_1s_2c_3|={\cal O}(1)$. The distributions in
Fig.~\ref{Fig:alphas-contlo} are seen to satisfy this condition.

While the allowed range of $M_{H^\pm}$ depends on the neutral Higgs boson mass
$M_2$, it is typically of the order of 300--700~GeV. In most of the allowed
parameter space, CP is violated, but along the edges $\alpha_2\to\pm\pi/2$
there is no CP-violation. This is the limit where the lightest Higgs boson,
$H_1$, becomes CP odd.

%%%%%%%%%%%%%%%%%%%%%%%%%%%%%%%%%%%%%%%%%%%%%%%%%%%%%%%%%%%%%%%%%%%%%%%%
\subsection{High values of $\tan\beta$} 
%%%%%%%%%%%%%%%%%%%%%%%%%%%%%%%%%%%%%%%%%%%%%%%%%%%%%%%%%%%%%%%%%%%%%%%%

\begin{figure}[htb]
\vspace*{-2mm}
\begin{center}
\includegraphics[width=95mm]{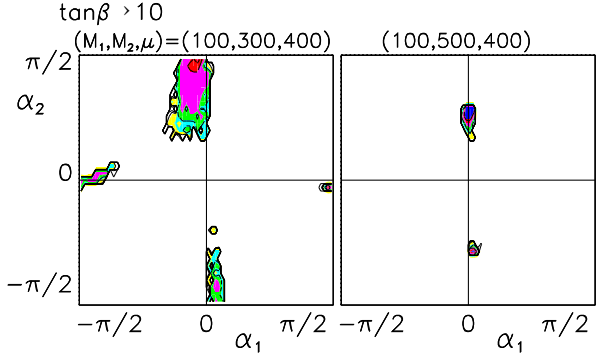}
\vspace*{-10mm}
\caption{\label{Fig:alphas-conthi} Normalized distributions of allowed regions
in the $\alpha_1$--$\alpha_2$ space, for high $\tan\beta$ values.  Contours
and colour codes are as in Fig.~\ref{Fig:alphas-contlo}, with additionally red
(blue) above $10^{-2}$ ($3\times10^{-2}$).}
\end{center}
\end{figure}

We show in Fig.~\ref{Fig:alphas-conthi} the populated regions in the
$\alpha_1$--$\alpha_2$ plane, for $\mu=400~\text{GeV}$ and $\tan\beta>10$.
Two points are worth noting: (1) The allowed regions satisfy the constraints
of (\ref{Eq:LEP2-cases}). (2) The majority of points do {\it not} satisfy the
condition (\ref{Eq:no-delta-rho}), meaning that the $M_{H^\pm}\gg M_1$ case is
not very relevant here. Instead, the degenerate case, $\mu\sim M_{H^\pm}\sim
M_2\lsim M_3$ is important, as also reflected in Table~\ref{tab:mh3-dist}~c).
%%%%%%%%%%%%%%%%%%%%%%%%%%%%%%%%%%%%%%%%%%%%%%%%%%%%%%%%%%%%%%%%%%%%%%
%%%%%%%%%%%%%%%%%%%%%%%%%%%%%%%%%%%%%%%%%%%%%%%%%%%%%%%%%%%%%%%%%%%%%%

At high values of $\tan\beta$, $\mu$ has to be comparable to $M_2$,
or higher. In particular, no solution exists for high values of $\tan\beta$
and $\mu=0$. This is due to the unitarity constraints.
Also, we note from Figs.~\ref{Fig:neu-all-mu=0}--\ref{Fig:neu-all-mu=600},
that unless $\mu$ is comparable to, or larger than $M_2$,
the allowed range in $\tan\beta$ and $M_{H^\pm}$ can be rather limited.
The range of
$M_{H^\pm}$ depends on the neutral Higgs boson mass $M_1$, being typically of
the order of 400--600~GeV.
In order to understand the large-$\tan\beta$ parameter space, let us review
\begin{align} \label{Eq:lambda1}
\lambda_1&=\frac{1}{c_\beta^2v^2}
[c_1^2c_2^2M_1^2
+(c_1s_2s_3+s_1c_3)^2M_2^2 \nonumber \\
&+(c_1s_2c_3-s_1s_3)^2M_3^2
-s_\beta^2\mu^2]>0.
\end{align}
This should be positive, but not ``too large''.
High values of $\mu$ require high $M_3$ and $|c_1s_2c_3-s_1s_3|={\cal O}(1)$.

%%%%%%%%%%%%%%%%%%%%%%%%%%%%%%%%%%%%%%%%%%%%%%%%%%%%%%%%%%%%%%%%%%%%%%%%
\subsection{Distribution of $M_3$}
%%%%%%%%%%%%%%%%%%%%%%%%%%%%%%%%%%%%%%%%%%%%%%%%%%%%%%%%%%%%%%%%%%%%%%%%

We recall that with our choice of parameters, the third neutral Higgs mass is
a derived quantity. The distribution of $M_3$ can be discussed
in terms of the dimensionless ratio
\begin{equation}
\xi=\frac{M_3}{M_2}, \quad
1<\xi.
\end{equation}
For three bins in $\xi$,
this is for $\mu=0$, 200, 400 and $600~\text{GeV}$
distributed as given in Table~\ref{tab:mh3-dist}.
For $\mu=0$, $M_3$ tends to be low, just marginally above $M_2$,
as seen in Table~\ref{tab:mh3-dist}~a).
This pattern is valid also for $M_1<\mu$, provided only that
$\mu<M_2$, see Table~\ref{tab:mh3-dist}~b) and c).
For $M_2<\mu$, on the other hand, $M_3$ can be large.
%%%%%%%%%%%%%%%%%%%%%%%%%%%%%%%%%%%%%%%%%%%%%%%%%%%%%%%%%%%%%%%%%%%%%%
\begin{table}[htb]
\begin{center}
\renewcommand{\tabcolsep}{.75em}
a)\quad $\mu=0$, $(M_1,M_2)=(100,300\ [500])$~GeV
\begin{tabular}{|c|c|c|c|}
\hline 
$\tan\beta$&$\xi<1.1$&$1.1<\xi<1.5$&$1.5<\xi$\\
\hline
$5-10$&94.6\ [\ 0.0]\%&5.2\ [ 0.0]\%&0.2\ [\ 0.0]\%\\
$<5$&43.4\ [88.4]\%&49.8\ [11.6]\%&6.8\ [\ 0.0]\%\\
\hline
\end{tabular}
\\
\vspace*{5mm}
b)\quad $\mu=200~\text{GeV}$, $(M_1,M_2)=(100,300\ [500])$~GeV
\begin{tabular}{|c|c|c|c|}
\hline 
$\tan\beta$&$\xi<1.1$&$1.1<\xi<1.5$&$1.5<\xi$\\
\hline
$>10$&74.0\ [95.8]\%&24.9\ [\ 4.2]\%&\ 1.0\ [\ 0.0]\%\\
$5-10$&49.0\ [91.4]\%&48.4\ [ 8.6]\%&\ 2.6\ [\ 0.0]\%\\
$<5$&30.9\ [81.5]\%&56.8\ [18.5]\%&12.3\ [\ 0.0]\%\\
\hline
\end{tabular}
\\
\vspace*{5mm}
c)\quad $\mu=400~\text{GeV}$, $(M_1,M_2)=(100,300\ [500])$~GeV
\begin{tabular}{|c|c|c|c|}
\hline 
$\tan\beta$&$\xi<1.1$&$1.1<\xi<1.5$&$1.5<\xi$\\
\hline
$>10$&0.0\ [91.6]\%&69.0\ [\ 8.4]\%&31.0\ [\ 0.0]\%\\
$5-10$&0.0\ [76.3]\%&47.8\ [23.7]\%&52.2\ [\ 0.0]\%\\
$<5$&0.0\ [64.1]\%&15.3\ [35.8]\%&84.7\ [\ 0.2]\%\\
\hline
\end{tabular} 
\\
\vspace*{5mm}
d)\quad $\mu=600~\text{GeV}$, $(M_1,M_2)=(100,300\ [500])$~GeV
\begin{tabular}{|c|c|c|c|}
\hline 
$\tan\beta$&$\xi<1.1$&$1.1<\xi<1.5$&$1.5<\xi$\\
\hline
$>10$&0.0\ [\ 0.0]\%&0.0\ [93.8]\%&100.0\ [\ 6.2]\%\\
$5-10$&0.0\ [\ 0.0]\%&0.0\ [92.6]\%&100.0\ [\ 7.5]\%\\
$<5$&0.0\ [\ 0.0]\%&0.0\ [94.5]\%&100.0\ [\ 5.6]\%\\
\hline
\end{tabular} 
\caption{Distribution of $M_3$ values, $\xi=M_3/M_2$.}
\label{tab:mh3-dist}
\end{center}
\vspace*{-5mm}
\end{table}
%%%%%%%%%%%%%%%%%%%%%%%%%%%%%%%%%%%%%%%%%%%%%%%%%%%%%%%%%%%%%%%%%%%%%%

%%%%%%%%%%%%%%%%%%%%%%%%%%%%%%%%%%%%%%%%%%%%%%%%%%%%%%%%%%%%%%%%%%%%%%%%
\subsection{The Standard-Model-like limit} 
%%%%%%%%%%%%%%%%%%%%%%%%%%%%%%%%%%%%%%%%%%%%%%%%%%%%%%%%%%%%%%%%%%%%%%%%
The parameters of the 2HDM can be chosen such that the $ZZH_1$,
$bbH_1$ and $ttH_1$ couplings all approach the corresponding
SM values. This requires, in our notation \cite{ElKaffas:2006nt}
\begin{align}
\cos(\beta-\alpha_1)\cos\alpha_2&\simeq 1, \nonumber \\
\frac{\cos\alpha_1\cos\alpha_2}{\cos\beta}&\simeq1, \nonumber \\
\frac{\sin\alpha_1\cos\alpha_2}{\sin\beta}&\simeq1,
\end{align}
which is satisfied for $\beta\simeq\alpha_1$ and $\alpha_2\simeq0$.  There
could also be a ``quasi-SM-like'' limit, where one or more of the above
quantities approaches $-1$ (denoted ``Solution B'' in \cite{Ginzburg:2001ss}).
For the familiar observables, such a sign change would not have any effect.

For $\tan\beta\lsim5$,
the allowed regions in  the $\alpha_1$--$\alpha_2$ space are rather extended,
and SM-like solutions are found for a range of mass values.
For $\tan\beta\gsim10$, on the other hand,
the populated parts of the $\alpha_1$--$\alpha_2$ space
become very localized, and have the following features:
(i) for $M_2\gsim\mu$, $\alpha_1\simeq0$ and $|\alpha_2|>0$, and 
(ii) for $M_2\lsim\mu$, additional regions emerge for small
values of $|\alpha_2|$ and $|\alpha_1|\simeq\pi/2$. The latter,
seen as small specks near the horizontal axis in the
left panel of Fig.~\ref{Fig:alphas-conthi},  correspond
to the SM-like (and ``quasi-SM-like'') case.

For the case $M_1=100~\text{GeV}$, which is studied
in most of our figures, there is of course a limit
to how close we come to the SM limit, since an SM Higgs mass of this 
low value is excluded. Actually, the fact that 
for $(M_1,M_2)=(100,500)~\text{GeV}$ and $\mu=0$, some region of
the $\tan\beta$--$M_{H^\pm}$ plane is excluded by the LEP2
non-discovery constraint (see right panel of Fig.~\ref{Fig:lep2-mu=200}),
means that those regions correspond to solutions near
the SM limit.

%%%%%%%%%%%%%%%%%%%%%%%%%%%%%%%%%%%%%%%%%%%%%%%%%%%%%%%%%%%%%%%%%%%%%%%%
\section{Possible future constraints} 
\label{sect:future}
\setcounter{equation}{0}
%%%%%%%%%%%%%%%%%%%%%%%%%%%%%%%%%%%%%%%%%%%%%%%%%%%%%%%%%%%%%%%%%%%%%%%%

\begin{figure}[htb]
\vspace*{-2mm}
\begin{center}
\includegraphics[width=95mm]{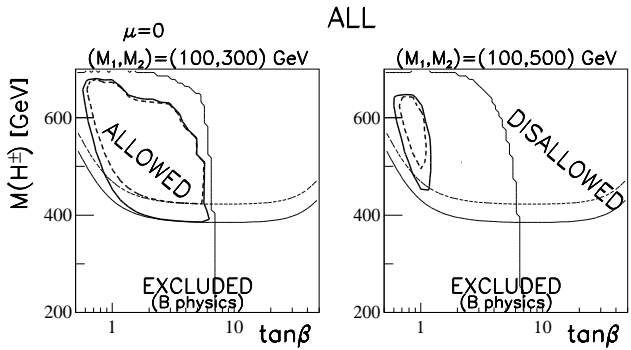}
\vspace*{-10mm}
\caption{\label{Fig:neu-all-mu=0-3.20} Similar to Fig.~\ref{Fig:neu-all-mu=0},
for $BR(\bar B\to X_s\gamma)=3.20\times10^{-4}$ and an over-all uncertainty
$\sigma[{\cal B}(\bar B\to X_s\gamma)]=0.25\times10^{-4}$
(c.f.\ Eq.~(\ref{Eq:bsgamma-sigma})).}
\end{center}
\end{figure}

\begin{figure}[htb]
\vspace*{-2mm}
\begin{center}
\includegraphics[width=95mm]{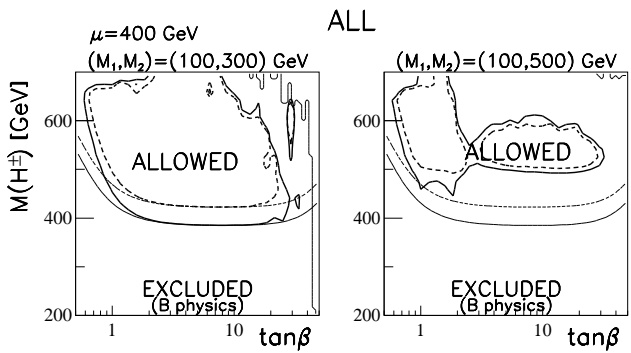}
\vspace*{-10mm}
\caption{\label{Fig:neu-all-mu=400-3.20} Similar to
Fig.~\ref{Fig:neu-all-mu=0-3.20}, for $\mu=400~\text{GeV}$.}
\end{center}
\end{figure}

It is interesting to see how the parameter-space constraints would be modified
by possible future results from the $B$-physics sector.  We shall not consider
any change to the $\Delta\rho$ or LEP2 constraints, but rather discuss the
possibility that the central value for the branching ratio for $\bar B\to
X_s\gamma$ be reduced from $3.55\times10^{-4}$ to $3.20\times10^{-4}$, a value
closer to the SM prediction. Also changing the overall uncertainty to
$0.25\times10^{-4}$, we find that the resulting constraints are significantly
modified, as illustrated in Figs.~\ref{Fig:neu-all-mu=0-3.20} and \ref{Fig:neu-all-mu=400-3.20}
for $\mu=0$ and 400~GeV, respectively.
%%%%%%%%%%%%%%%%%%%%%%%%%%%%%%%%%%%%%%%%%%%%%%%%%%%%%%%%%%%%%%%%%%%%%%
%%%%%%%%%%%%%%%%%%%%%%%%%%%%%%%%%%%%%%%%%%%%%%%%%%%%%%%%%%%%%%%%%%%%%%
%%%%%%%%%%%%%%%%%%%%%%%%%%%%%%%%%%%%%%%%%%%%%%%%%%%%%%%%%%%%%%%%%%%%%%
%%%%%%%%%%%%%%%%%%%%%%%%%%%%%%%%%%%%%%%%%%%%%%%%%%%%%%%%%%%%%%%%%%%%%%

%%%%%%%%%%%%%%%%%%%%%%%%%%%%%%%%%%%%%%%%%%%%%%%%%%%%%%%%%%%%%%%%%%%%%%%%
\section{Summary} 
\label{sect:summary}
\setcounter{equation}{0}
%%%%%%%%%%%%%%%%%%%%%%%%%%%%%%%%%%%%%%%%%%%%%%%%%%%%%%%%%%%%%%%%%%%%%%%%
We have shown that the $B$-physics results, together with the precise
measurement of the $\rho$-parameter at LEP and the constraint of tree-level
unitarity of Higgs--Higgs scattering, exclude large regions of the 2HDM~(II)
parameter space.  High values of $\tan\beta$ are excluded unless both $M_2$
and $M_3$ are heavy. Furthermore, they should be reasonably close to each
other.  Improved precision of the $\bar B\to X_s\gamma$ measurement could
significantly reduce the remaining part of the parameter space, but it appears
unlikely that the model could be excluded other than by a negative search at
the LHC.

What is the corresponding situation for supersymmetric models?  While the
consistency then is guaranteed by internal relations, it should be kept in
mind that light charged Higgs bosons would be in conflict with the $B$-physics
data unless some superpartner (for example, the chargino \cite{Bobeth:1999ww})
is also light.  A possibility which has received some attention, is a light
chargino and a light stop \cite{Barbieri:1993av}. It has also been shown that
anomalous effects at large $\tan\beta$ could weaken the bound on $M_{H^\pm}$
without light superpartners \cite{Degrassi:2000qf} (see also
\cite{Carena:2000uj}).  However, the more recent data on $B\to\tau\nu_\tau$
discussed in Sec.~\ref{Sec:B-tau-nu} would presumably close this loophole
(see Fig.~2 of \cite{Haisch:2007ic}).

Similar scans over the parameter space of the Constrained MSSM
\cite{Ellis:2006ix,Roszkowski:2007fd,Ellis:2007fu} differ from the present
work in one major respect: they are required to yield an amount of
dark matter that is compatible with the WMAP data
\cite{Spergel:2006hy}. 
Additionally, the $a_\mu$ constraint is more severe, due to
one-loop contributions involving superpartners of the muon and muon neutrino.
These studies are also more focused on the high-scale parameters, like $m_0$
and $m_{1/2}$, with less emphasis on $M_{H^\pm}$ and $\tan\beta$.  The study
by \cite{Roszkowski:2007fd} shows a preference for positive $\mu$ (higgs\-ino
mass parameter, not to be confused with the $\mu$ of Eq.~(\ref{Eq:musq})), a
relatively light charged Higgs mass (a few hundred GeV) and rather high values
of $\tan\beta$ ($\sim50-60$).

\bigskip

%%%%%%%%%%%%%%%%%%%%%%%%%%%%%%%%%%%%%%%%%%%%%%%%%%%%%%%%%%%%%%%%%%%%%%%
{\bf Acknowledgments.}  It is a pleasure to thank Mikolaj Misiak for patiently
explaining details of the $\bar B\to X_s\gamma$ calculation, and Andrzej
Buras, Ulrich Jentschura and Frank Krauss, for correspondence related to
reference \cite{Urban:1997gw}.  This research has been supported in part by
the Mission Department of Egypt and the Research Council of Norway.
%%%%%%%%%%%%%%%%%%%%%%%%%%%%%%%%%%%%%%%%%%%%%%%%%%%%%%%%%%%%%%%%%%%%%%%%%%%%%%
%\clearpage

%%%%%%%%%%%%%%%%%%%%%%%%%%%%%%%%%%%%%%%%%%%%%%%%%%%%%%%%%%%%%%%%%%%%%%%%

\end{document}